\documentclass[english,aps,prl,amssymb,amsfont,twocolumn,showpacs,floatfix,superscriptaddress,nobalancelastpage,footinbibreprint]{revtex4-1}
\usepackage[latin9]{inputenc}
\usepackage{amsmath}
\usepackage{graphicx}
\usepackage{epstopdf}
\usepackage{url}
\begin{document}
\title{Tunable Superconducting Qubits with Flux-Independent Coherence}

\author{M. D. \surname{Hutchings}}
%\email{mdhutc01@syr.edu}
\affiliation{Syracuse University, Department of Physics, Syracuse, NY 13244, USA}

\author{J. B. \surname{Hertzberg}}

\affiliation{IBM, TJ Watson Research Center, Yorktown Heights, NY 10598, USA}

\author{Y. \surname{Liu}}

\affiliation{Syracuse University, Department of Physics, Syracuse, NY 13244, USA}

\author{N. T. \surname{Bronn}}

\affiliation{IBM, TJ Watson Research Center, Yorktown Heights, NY 10598, USA}

\author{G. A. \surname{Keefe}}

\affiliation{IBM, TJ Watson Research Center, Yorktown Heights, NY 10598, USA}

\author{J. M. \surname{Chow}}

\affiliation{IBM, TJ Watson Research Center, Yorktown Heights, NY 10598, USA}

\author{B. L. T. \surname{Plourde}}

\affiliation{Syracuse University, Department of Physics, Syracuse, NY 13244, USA}

\date{\today}

\begin{abstract}
%We have studied the impact of low-frequency magnetic flux noise upon superconducting transmon qubits %with differing levels of tunability. We find that qubits with weaker tunability exhibit dephasing that is less %sensitive to flux noise. This insight was used to fabricate qubits where dephasing due to flux noise was %suppressed below other dephasing sources, leading to flux-independent dephasing times $T_2^* \sim 15\,%\mu$s over a tunable range of $\sim340\,{\rm MHz}$. Tunable transmon qubits of this type could lower the %overhead required to reach fault-tolerant qubit gate operation and fundamentally improve scalability for the %creation of a quantum processor.
We have studied the impact of low-frequency magnetic flux noise upon superconducting transmon qubits with various levels of tunability. We find that qubits with weaker tunability exhibit dephasing that is less sensitive to flux noise. This insight was used to fabricate qubits where dephasing due to flux noise was suppressed below other dephasing sources, leading to flux-independent dephasing times $T_2^* \sim 15\,\mu$s over a tunable range of $\sim340\,{\rm MHz}$. Such tunable qubits have the potential to create high-fidelity, fault-tolerant qubit gates and fundamentally improve scalability for a quantum processor.
\end{abstract}

\keywords{list some}

\maketitle

Quantum computers have the potential to outperform classical logic in important technological problems. A practical quantum processor must be comprised of quantum bits (\textquotedblleft qubits") that are isolated from environmental decoherence sources yet easily addressable during logical gate operations. Superconducting qubits are an attractive candidate because of their simple integration with fast control and readout circuitry. In recent years, advances in superconducting qubits have demonstrated how such integration may be achieved while maintaining high coherence  \citep{barends2014superconducting,kelly2015state,corcoles2015demonstration}. Further extensions of qubit coherence will serve to reduce gate errors, cutting down the number of qubits required for fault-tolerant quantum logic \citep{Gambetta2017Build,fowler2012surface}.

%Quantum computers have the potential to outperform classical logic in several important technological
%problems. A scalable realization of a quantum processor calls for quantum bits (\textquotedblleft qubits") that are well isolated
%from the environment, yet easily addressed so that quantum logic gates can be
%implemented with high fidelity. These requirements are conflicting making an experimental realization of a quantum computer a challenge. Superconducting qubit are a leading candidate to overcome these challenges, their
%strong interactions with quantum circuits allows for simple integration with fast control and readout. However, this fundamentally impacts the ability to maintain the coherence of any qubit. Though substantial advances in qubit coherence have been achieved in recent years \citep{barends2014superconducting,kelly2015state,corcoles2015demonstration}, Reducing decoherence further would lower errors in current fault tolerant gates: cutting down qubit quantity overhead and 
%creating truly scalable quantum computer.  
An important aspect of maintaining high qubit coherence is the reduction of dephasing. Frequency-tunable qubits are inherently sensitive to dephasing via noise in the tuning control channel. Tuning via a magnetic flux thus introduces dephasing via low-frequency flux noise \citep{anton2012pure,yoshihara2006decoherence,bylander2011noise,ithier2005decoherence,paladino20141,martinis2003decoherence,wellstood1987low,quintana2017observation}. Such noise is ubiquitous in thin-film superconducting
devices at low temperatures. Experiments indicate a high density of unpaired spins on the thin-film surface \citep{sendelbach2008magnetism} with fluctuations of these leading to low-frequency flux noise that typically has a $1/f$ power spectrum \citep{faoro2008microscopic,wang2015candidate,laforest2015flux,quintana2017observation}. For any flux-tunable qubit, this flux noise leads to significant dephasing whenever the qubit is biased at a point with a large gradient of the qubit energy with respect to flux.

Flux tuning is nonetheless highly advantageous for many quantum circuits, and several classes of quantum logic gates rely on flux-tunable qubits. In the controlled-phase gate \citep{barends2014superconducting,dicarlo2009demonstration}, qubit pairs are rapidly tuned into resonance to create entanglement. Here, both flux noise and off-resonant coupling to other qubits produce phase errors proportional to gate times, with total gate error scaling as the square of the gate time \citep{martinis2014fast}. Alternatively, fixed-frequency qubits have been employed in schemes such as the cross resonance (CR) gate \citep{Rigetti2010fully,chow2011simple} to demonstrate aspects of quantum error correction (QEC) \citep{chow2014implementing,corcoles2015demonstration}. Recent efforts with two-qubit devices have extended CR gate fidelities beyond 99\% \citep{Sheldon_Procedure_2016}. Larger lattices of fixed-frequency qubits, however, are likely to suffer increasingly from frequency crowding. If a qubit's 0-1 excitation frequency overlaps with the 0-1 or 1-2 frequency of its neighbor, or if the two qubits' frequencies are very far apart, the CR gate between these two qubits will be non-ideal, with the strong possibility of leakage out of the computational subspace, or a very weak gate, respectively \citep{divincenzo2013quantum}. However, fixed-frequency transmon qubits are challenging to fabricate to precision better than about $200\,{\rm MHz}$ \citep{privcommsrosenblatt}. Given such imprecision, a hypothetical seventeen-qubit logic circuit could see up to a quarter of its gate pairs fail due to frequency crowding (see Supplement). Frequency-tunable transmon qubits therefore appear attractive for use in architectures based on the CR gate.

In this Letter, we show how a tunable qubit's sensitivity to flux noise may be reduced by limiting its extent of 
tunability. We report results for several different qubits showing that the qubit dephasing rate is proportional to the sensitivity of the qubit frequency to magnetic flux and to the amplitude of low-frequency flux noise.
%qubit dephasing is proportional to the 
%rate of change of qubit frequency with magnetic flux and to the strength of any coupled flux noise. 
Furthermore, we use the understanding gained through this study to fabricate a qubit whose dephasing due 
to non-flux dependent sources exceeds its dephasing due to low-frequency flux noise, over a range of more 
than 300 MHz of tunability. This unique qubit has the potential to reduce errors in gates employing 
frequency-tunable qubits and to evade frequency crowding in qubit lattices employing CR gates. It therefore 
offers a promising route to create high-fidelity two-qubit gates that reach fault-tolerant gate operation and to improve 
the scalability of superconducting qubit devices.

Our device adapts a design in which a superconducting quantum interference device (SQUID) serves as the 
Josephson inductance in a transmon qubit \citep{koch2007charge}. Here, the Josephson energy, and 
consequently the qubit 0-1 transition frequency $f_{01}$, may be tuned with a magnetic flux $\Phi$ with a period of $\Phi_0 \equiv h/2e$, the magnetic flux quantum, where $h$ is Planck's constant and $e$ is the electron charge. However, if the two junctions in the SQUID have different Josephson energies $E_{J1}$ and $E_{J2}$, a 
so-called `asymmetric transmon' is formed \citep{strand2013first}. The greater the difference in junction 
energies, the smaller the level of tunability. 
If $E_{J1}$~\textgreater~$E_{J2}$, we can define the ratio $\alpha = E_{J1}/E_{J2}$ and the sum $E_{J
\Sigma} = E_{J1} + E_{J2}$. The total flux-dependent Josephson energy $E_J$ varies according to the 
following expression from Ref. \citep{koch2007charge}:

%\begin{equation}
%E_{J}(\varPhi)=E_{j\Sigma}\cos\left(\phi\right)\sqrt{1+d^{2}\tan^{2}\left(\phi\right)}\label{eq:1},
%\end{equation}
%
\begin{equation}
E_{J}(\mathrm{\Phi})=E_{J\Sigma}\cos{\bigg( \frac{\pi \Phi}{\Phi_{0}}}\bigg)\sqrt{1+d^2\tan^{2}{\bigg(\frac{\pi \Phi}{\Phi_{0}}\bigg)}}
\label{eq:1},
\end{equation}
where $d$ is given by $d=(\alpha-1)/(1+\alpha)$.  

%In this Letter, we show how the sensitivity to flux
%noise for a tunable qubit can be reduced by limiting its range of tunability. Flux-tunable qubits are created by integrating a superconducting quantum interference device (SQUID) into the Josephson junction geometry \citep{koch2007charge}. The dc SQUID allows the Josephson energy, and consequently the qubit frequency $f_{01}$, to be tuned periodically with magnetic flux $\Phi$. By designing
%the two junctions in the dc SQUID to have different junction energies,
%the level of tunability can be adjusted; this creates an asymmetric
%transmon \citep{strand2013first}. The greater the difference in junction
%energies, the smaller the level of tunability. We report results for
%qubits with a range of tunabilities showing that qubit dephasing is
%proportional to the rate of change of qubit frequency with magnetic
%flux and the strength of any coupled flux noise. Furthermore, we use the understanding gained through 
%this study to fabricate a qubit where the level of dephasing due to low-frequency flux noise has been 
%suppressed below the background set by other non-flux dependent dephasing sources. This leading to flux-independent coherence over more than 300 MHz of tunability. This unique qubit has the potential to reduce errors in gates employing frequency tunable qubits and minimize the challenges facing fixed-frequency qubit gates. This will lower the overhead required to reach fault tolerant gate operation and fundamentally improving the scalability of superconducting qubits devices.

To explore the dephasing behavior of qubits having such tunability, we prepared transmon qubits on two styles of chip, referred
to as sample A and sample B. Both samples employ a multi-qubit planar circuit quantum electodynamics (cQED) architecture with eight separate cavity/qubit systems. Qubits on the same chip should experience the same flux noise level, allowing a comparison of dephasing properties between them. On sample A, the eight frequency-multiplexed cavities are all coupled to a common feedline for microwave drive and readout (sample details and layout shown in Supplement). On this chip, we compare transmons having junction ratios $\alpha = 7$, 4, 1 and a fixed-frequency single-junction qubit. We design for a particular $\alpha$ by varying the junction areas in the SQUID, since $E_{J}$ of a junction is directly proportional to its area. For consistency, the single-junction qubit maintained the same SQUID loop structure with one of the junctions being left open, and all four qubit types were designed to have the same $E_{J\Sigma}$. Sample B employs a qubit design similar to that in \citep{chow2014implementing,corcoles2015demonstration,Takita2016Dem,ibmquantumexp}. For this device, all qubits have separate drive and readout microwave lines (layout shown in Supplement). Six qubits were designed to have $\alpha = 15$ while two employed a single junction matching the $E_{J1}$ of the tunable qubits. The fixed-frequency qubits act as a reference for non-flux dependent dephasing on each chip.

\begin{figure}
\includegraphics[width=8.6cm]{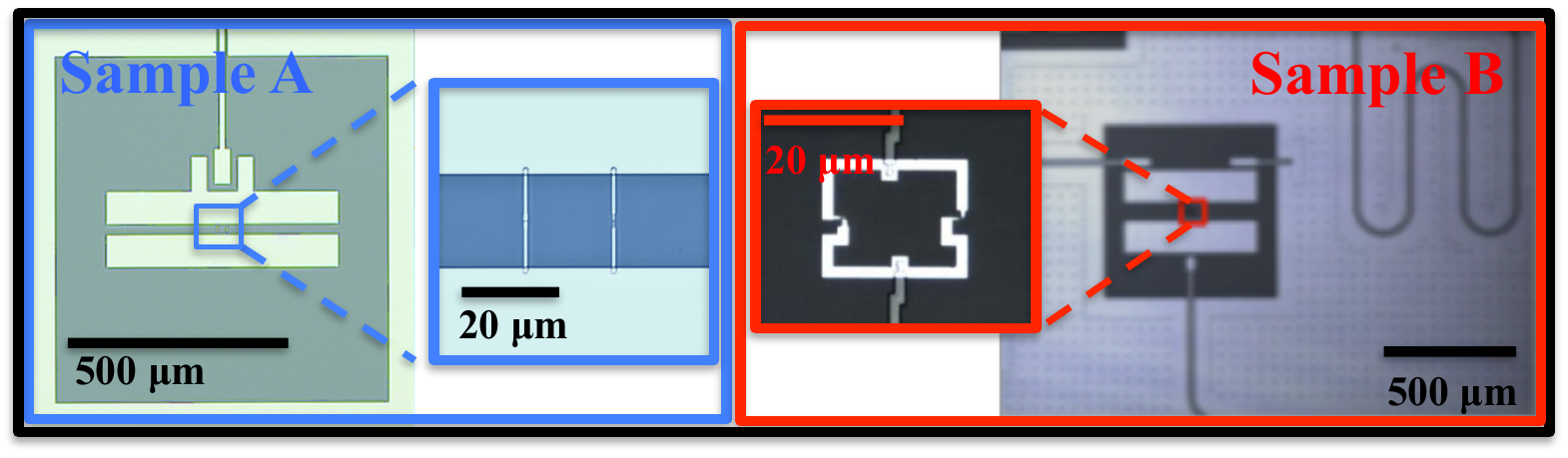}
\caption{(color online) Optical micrographs of example qubits from samples A and B.
\label{FIG 1}}
\end{figure}

We used standard photolithographic and etch processes to pattern the coplanar waveguides, ground plane, and qubit capacitors from Nb sputtered films on Si substrates, followed by electron-beam lithography and deposition of conventional Al-AlO$_{x}$-Al shadow-evaporated junctions. While all qubits were similar in design to \citep{Sheldon_Procedure_2016,chow2014implementing,corcoles2015demonstration}, the transmon capacitor pads and SQUID loop geometry differed somewhat between samples A and B. Designs are shown in Fig. \ref{FIG 1}. Each sample was mounted on a dilution refrigerator in its respective lab (sample A at Syracuse; sample B at IBM) and surrounded by both room-temperature and cryogenic magnetic shields. Measurements for both samples were performed using standard cQED readout techniques \citep{Reed_High_2010}. Measurement signals from both samples were amplified by a low-noise HEMT amplifier at 4K. In the case of the $\alpha = 15$ qubit on sample B, additional amplification was provided by a SLUG amplifier \citep{HoverAPL2014_104_152601}. Flux bias was applied to each sample during measurement using a wire coil placed close to the top of each device. Fabrication details and a discussion of measurement techniques are given in the Supplement.

%Sample A was designed to have four qubit variations with two copies
%of each qubit. These included an asymmetric
%7:1 qubit, with one junction designed to have a Josephson energy seven times larger than the other, a 4:1 qubit, a symmetric
%1:1 qubit and a single junction fixed-frequency qubit. It was important to have a consistent design between qubits so the single %junction qubit maintained the same SQUID loop structure with one of
%the electrodes being left open. This fixed-frequency qubit was included so that non-flux dependent background dephasing sources could be compared between qubits. We define the ratio of the two junctions Josephson energy as $\alpha$. All qubits on sample A were designed to have the same total Josephson energy, thus they should have similar maximum qubit frequencies. Sample B had two qubit types, a single-junction fixed-frequency qubit and a 15:1 asymmetric transmon, with four copies of each on a single chip. The qubits on sample B were also designed to have the same
%maximum qubit frequency across the chip, as in sample A. 

\begin{figure}
\includegraphics[width=8.6cm]{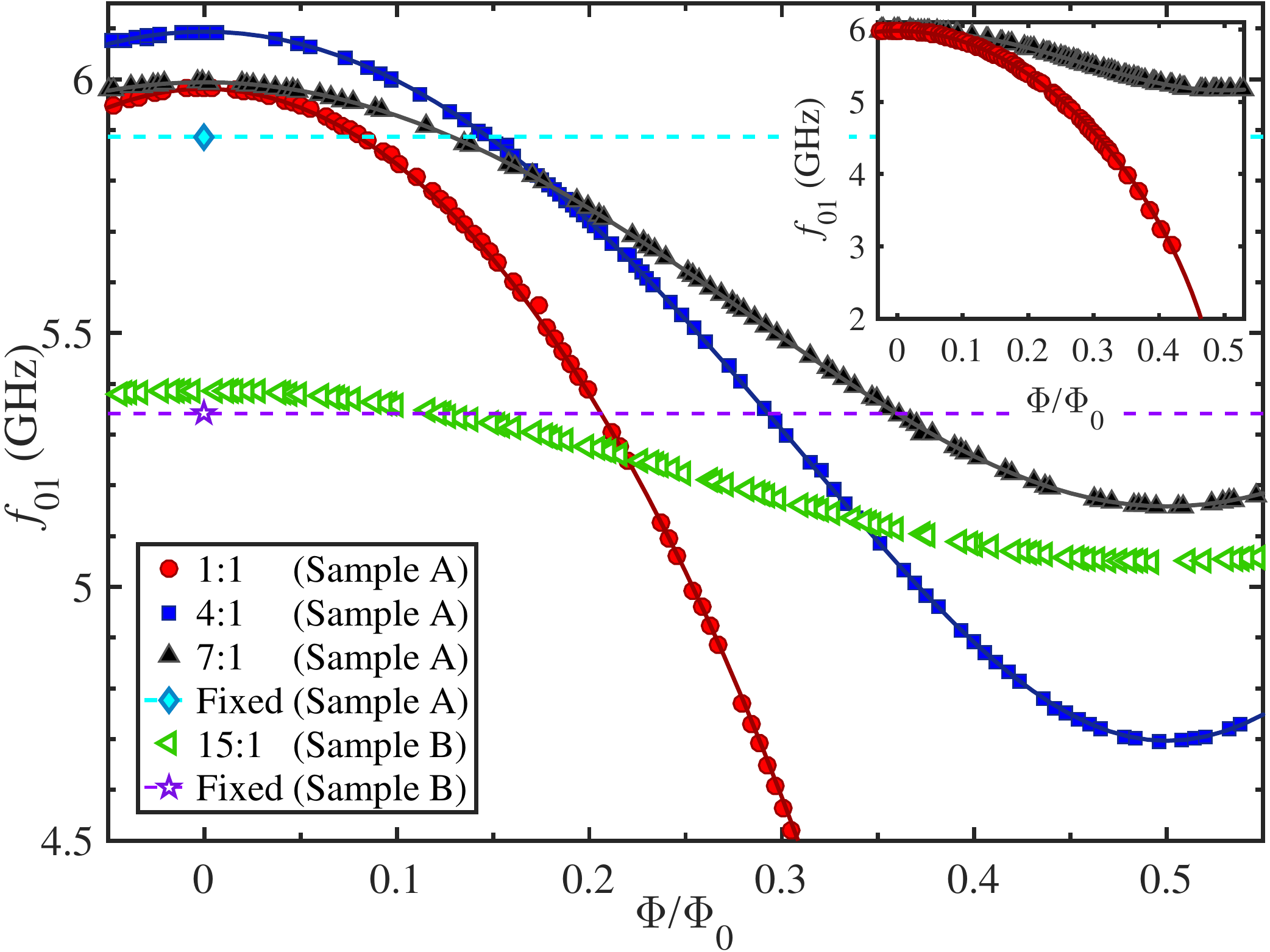}
\caption{(color online) $f_{01}$ vs. flux measured for qubits from samples A and B. Solid lines are fits to these tuning curves based on Eq. 1. Also included are frequencies of single junction qubits from both samples. Dashed lines for these qubits to guide the eye. Inset: entire tuning range measured for the $\alpha =1$ qubit with the $\alpha =7$ qubit included as a comparison to illustrate the large frequency tunability of an $\alpha =1$ qubit.
\label{FIG 2}}
\end{figure}

Here we present data from four qubits on sample A and two qubits on sample B, one of each variation from each sample. Figure \ref{FIG 2}  shows the flux dependence of $f_{01}$ for each qubit. We have subtracted any fixed flux offset appearing in the measurement. The $\alpha = 15$ qubit on sample B had the weakest tunability of these: $337\,{\rm MHz}$. Following Eq. {(}\ref{eq:1}{)} and the expectation that $f_{01} \propto \sqrt{E_J}$ \citep{koch2007charge}, we fit the data in Fig. \ref{FIG 2} to find the maximum frequency $f_{01}^{\mathrm {max}} \propto \sqrt{E_{J\Sigma}}$ and asymmetry parameter $d$.
%\begin{align}
%f_{01} = f_{01}^0 [\cos^2(\phi) + d^2 \sin^2(\phi)]^{1/4} \label{Eq:f01}
%\end{align}
From the latter we compute $\alpha$ for all tunable qubits and we find that the measured asymmetry $\alpha$ was within 5\% of the designed value. We note that the four sample A qubits shown in Fig. \ref{FIG 2} were designed to have identical $E_{J\Sigma}$ and therefore identical $f_{01}^{\mathrm {max}}$, but in fact exhibit a $\sim 200\,{\rm MHz}$ spread, thus illustrating the challenge of fabricating qubits to precise frequencies.

To assess the effect of flux noise on dephasing, we observe how the latter relates to each qubit's frequency gradient as a function of flux $D_{\Phi} = |\partial{f}_{01}/\partial\Phi|$. We characterize dephasing via measurement of the Ramsey decay time $T_{2}^{*}$, which is sensitive to low-frequency dephasing noise \citep{yoshihara2006decoherence,ithier2005decoherence}. We fit these using an exponential form. Although it has been shown that a dephasing noise source with a $1/f$ power spectrum will result in a Gaussian decay envelope \citep{ithier2005decoherence,yoshihara2006decoherence}, flux-independent dephasing sources such as cavity-photon shot-noise \citep{sears2012photon,schuster2005ac,gambetta2006qubit} result in an exponential decay envelope. Ramsey decays for fixed-frequency qubits are therefore well fit with an exponential decay envelope. For all of our asymmetric transmons, as well as a large portion of the dephasing data for the $\alpha = 1$ symmetric device, we find that an exponential decay envelope is also a good fit.
%In all of our data the $T_{2}^{*}$ value found from an exponential fit exhibits a maximum difference of  {[}\%{]} compared to the value derived from a Gaussian expression. 
In all of our data, we find that differences between values of $T_{2}^{*}$ obtained using an exponential or Gaussian fit are systematic but slight. Furthermore, assuming a purely exponential decay simply puts an upper bound on the extracted flux noise level. A more complete discussion of the nature of our Ramsey decay envelopes and alternative fitting approaches appears in the Supplement.

Relaxation times $T_{1}$ ranged from $\sim 20 - 50\,\mu{\rm s}$ over the six qubits reported here. In general $T_{1}$ increased with decreasing qubit frequency {(}Supplement, Fig. 2{)}, consistent with dielectric loss and a frequency-independent loss tangent, as observed in other tunable superconducting qubits \citep{barends2013coherent}. For the $\alpha =15$ qubit on sample B, a reduction in $T_{1}$ with increasing frequency is also consistent with Purcell losses to the readout resonator. Qubits on sample A remained sufficiently detuned below the readout resonators that Purcell loss was not a significant loss channel.
$T_{1}$ relaxation due to coupling to a flux-bias line, first discussed for inductive coupling in Ref.  \citep{koch2007charge} for a near symmetrical qubit, and for capacitive coupling in Ref.  \cite{JohnsonDisserationYale2011}, was considered for the qubits studied here. We show in the Supplement that the upper bound on $T_{1}$ due to the flux-line coupling for our qubit designs is not significantly lower than that reported in \citep{koch2007charge}. 

\begin{figure}
\includegraphics[width=8.6cm]{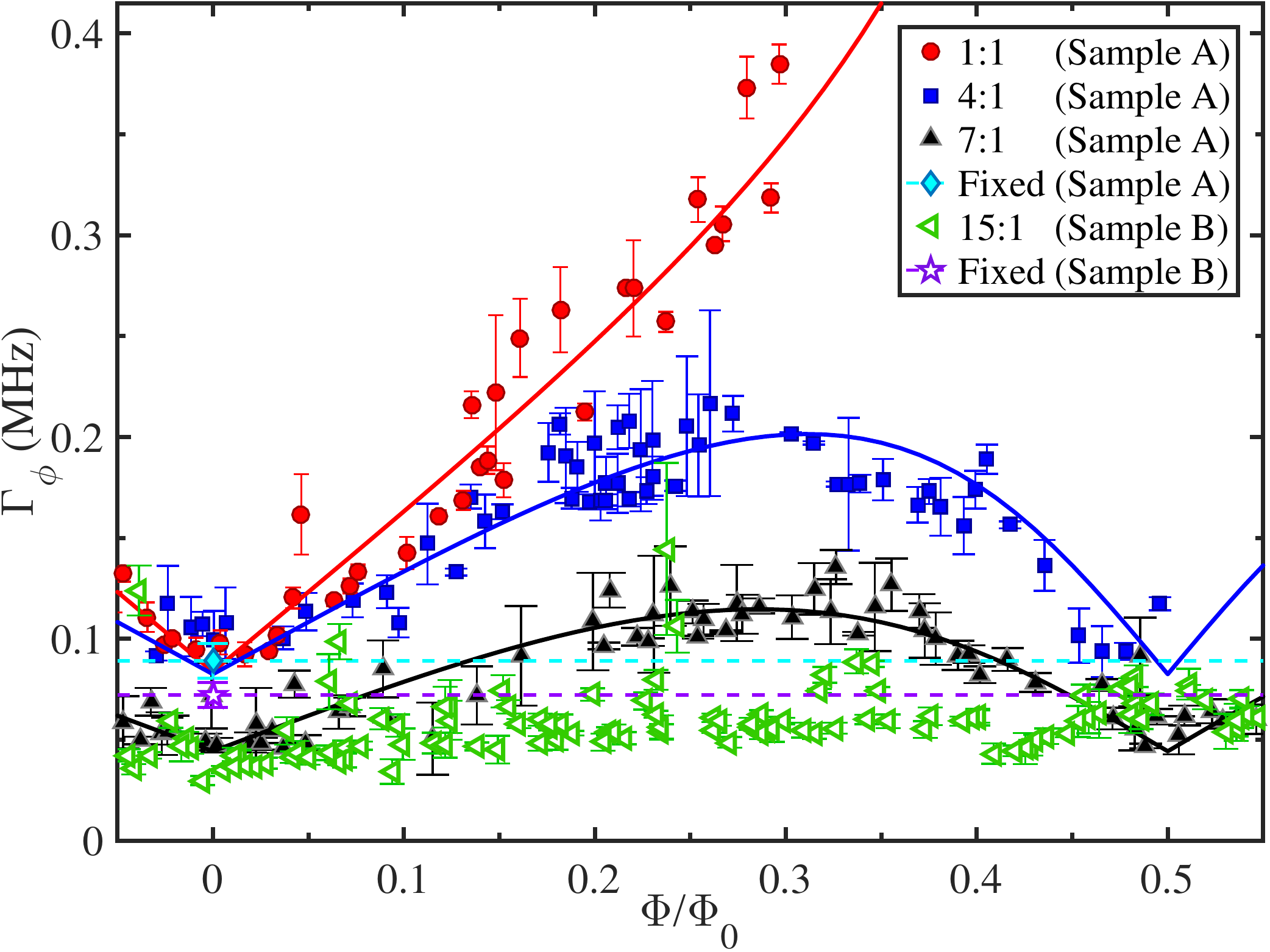}
\caption{(color online) $\Gamma_{\phi}$ vs. flux measured for qubits from samples A and B. Solid lines show a simultaneous fit of the form $mD_{\Phi} + b$ to the tunable qubits on sample A. Factor $m$ is common to all three datasets while $b$ is allowed to vary for each. $\Gamma_{\phi}$ measured for fixed frequency qubits on both samples included with dashed lines to help guide the eye.
\label{FIG 3}}
\end{figure}

To compare dephasing rates among the qubits, we use the relation $\Gamma_{\phi}=1/T_{2}^{*}-1/2T_{1}$ \citep{makhlin2001quantum} to remove the relaxation contribution. These values are plotted against flux in Fig. \ref{FIG 3}. As the curves in Fig. \ref{FIG 2} illustrate, the integer and half-integer $\Phi/\Phi_0$ points are `sweet spots' where $D_{\Phi} = 0$ and thus the qubit is first-order insensitive to flux noise. All the transmons on sample A clearly exhibit a dephasing rate that increases with $D_{\Phi}$ and is a minimum at the sweet spots. Second-order sensitivity to flux noise \citep{makhlin2004dephasing,ithier2005decoherence} should be negligible in our samples because of the small energy-band curvature. However, the level of $\Gamma_{\phi}$ for the non-tunable qubit on each sample and the tunable qubits at their sweet spots indicates the presence of non-flux dependent sources of dephasing. Such background dephasing may arise from other mechanisms, including cavity-photon shot noise \citep{sears2012photon}, critical current noise \citep{van2004decoherence}, or charge noise affecting the residual charge dispersion in the transmon design \citep{koch2007charge}. This background dephasing may be expected to vary
%
%However, non-flux dependent sources of dephasing such as cavity-photon shot noise and critical-current %noise \citep{van2004decoherence} are not negligible, as indicated by the $\Gamma_{\Phi}$ of the non-%tunable qubit on each sample. Such `background' dephasing may be expected to
%
from qubit to qubit due to differences in qubit-cavity coupling or cavity thermalization, among other effects. Such variations are commonly observed in multi-qubit devices \citep{corcoles2015demonstration,chow2014implementing,Takita2016Dem}. The Supplement contains dephasing data for additional devices similar to those discussed here, illustrating further variations in background dephasing.

For sample A, if we consider only flux-dependent dephasing,
%For sample A, if such background dephasing levels are neglected, 
it is evident that $\Gamma_{\phi} \propto D_{\Phi}$. Furthermore, qubits of the same geometry on the same chip should experience similar flux noise \citep{sendelbach2008magnetism}. The analysis outlined in Ref. \citep{ithier2005decoherence,yoshihara2006decoherence} may then be used to extract a flux noise level from the relationship between $\Gamma_{\phi}$ and $D_{\Phi}$. We apply a simultaneous fit of the form $m D_{\Phi} + b$ to the $\alpha = 1$, 4, and 7 qubits, allowing background dephasing $b$ to vary for each qubit, while a single $m$ is common to all.
%with $m$ being fixed between qubits and $b$ accounting for a background dephasing rate, which may vary from qubit to qubit. 
The fit appears as solid lines in Fig. \ref{FIG 3}. We derive $\Gamma_{\phi}=2\pi\sqrt{A_{\Phi}|\ln{(2\pi f_{IR}t)}|}D_{\Phi}$ following the approach in Ref. \citep{ithier2005decoherence}, where the flux noise power spectrum is $S_{\Phi}(f) = A_{\Phi}/|f|$, $f_{IR}$ is the infrared cutoff frequency, taken to be $1\,{\rm Hz}$ and $t$ is on the order of $1/\Gamma_{\phi}$, which we take to be $10\,\mu {\rm s}$ in our calculations. 
%Using this equation, we can calculate the flux noise level impacting the qubits on sample A from the simultaneous fit parameter $m$.
Equating $mD_{\Phi}$ to $\Gamma_{\phi}$ in the equation above, we may calculate the flux noise level on sample A.
To determine the uncertainty in the measured flux noise level, we must not only account for the error in fitting $m$ but also how variations in dephasing time impact the calculation of $A_{\Phi}$ values. To account for the latter, we determine the impact on extracted $A_{\Phi}^{1/2}$ as $t$ is varied. Adjusting $t$ over a range similar to what we observe experimentally leads to a $\sim10\%$ change $A_{\Phi}^{1/2}$. The errors we report for all calculated $A_{\Phi}^{1/2}$ reflect this added uncertainty. We find that $A_{\Phi}^{1/2} = 1.4 \pm 0.2$~$\mu \Phi_0$ on sample A. This level is compatible with previous experimental studies of flux noise in superconducting flux \citep{anton2012pure,yoshihara2006decoherence,bylander2011noise,orgiazzi2016flux,stern2014flux} and phase qubits \citep{bialczak2007flux}.  

%Figure \ref{FIG 3} shows the rate of dephasing $\Gamma_{\phi}$ measured for each qubit, calculated at each flux point using the standard relation: $\Gamma_{\phi}=1/T_{2}^{*}-1/2T_{1}$ \citep{makhlin2001quantum} to subtract off the relaxation contribution.
%The fits to Eq. \ref{eq:1}, shown in Fig.\ref{FIG 2},
%are used to scale DC voltage to magnetic flux level. 
%It is clear that $\Gamma_{\phi}$
%increases for the $\alpha = 7$ and less asymmetric qubits away from the integer and odd half-integer flux quantum bias points. These correspond to the so-called
%sweet spots of the qubits where $D_{\Phi} = 0$ and thus the qubit is first-order insensitive to flux noise. Although second-order sensitivity to flux noise has been considered \citep{makhlin2004dephasing,ithier2005decoherence} this is negligible in our samples because of the small energy-band curvature and background-dephasing processes dominate. The fixed-frequency qubits, not impacted by flux
%noise, provided comparisons of background dephasing rates and insight into non-flux dependent sources of dephasing. Such sources include cavity-photon shot noise and critical current noise \citep{van2004decoherence}.

\begin{figure}
\includegraphics[width=8.6cm]{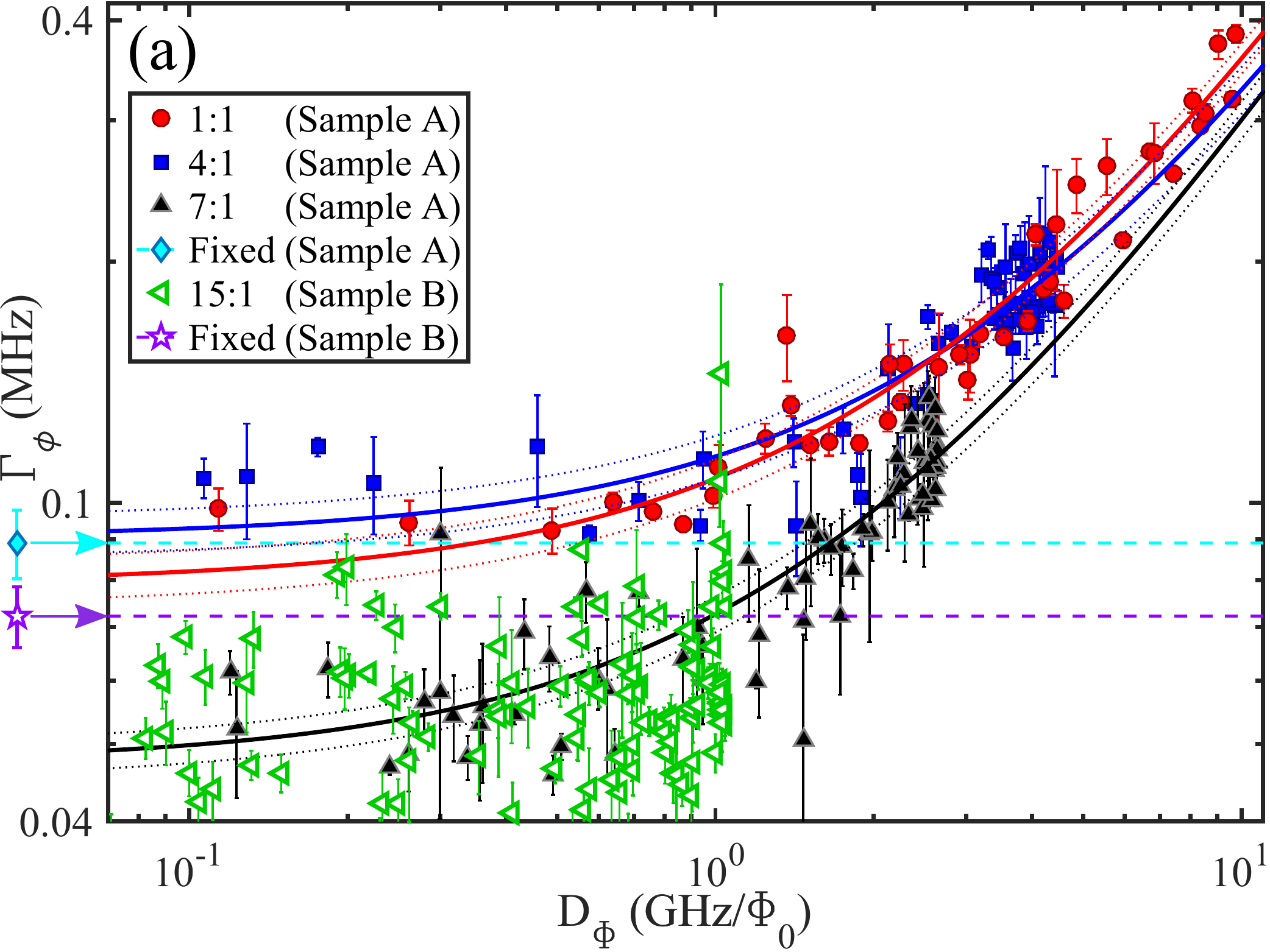}
\includegraphics[width=8.6cm]{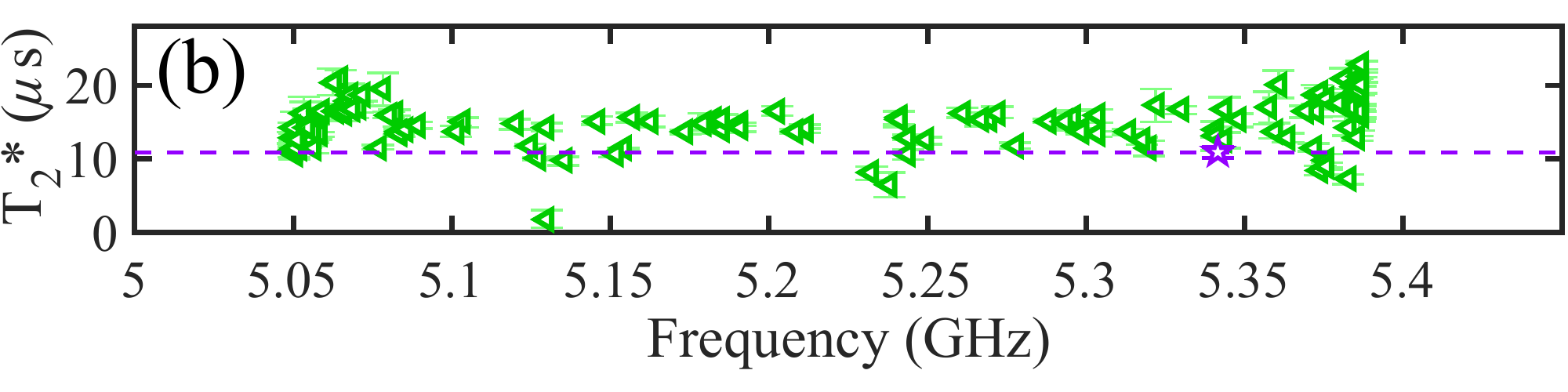}
\caption{(color online) {(}a{)} $\Gamma_{\phi}$ vs. $D_{\Phi}$ measured for qubits from samples A and B. Solid lines show individual linear fits to the tunable qubits on sample A, as described in text. Note the scale is log-log. $\Gamma_{\phi}$ measured for fixed-frequency qubits on both samples, included with dashed lines to help guide the eye. {(}b{)} $T^{*}_{2}$ vs. frequency measured for the $\alpha = 15$ and fixed-frequency qubits on sample B. 
\label{FIG 4}}
\end{figure}

To achieve an even clearer picture of the influence of flux noise on these qubits, we plot $\Gamma_{\phi}$ vs. $D_{\Phi}$ for each qubit in Fig. \ref{FIG 4}a. Here, $D_{\Phi}$ is computed from the fits to the energy bands of each qubit shown in Fig. \ref{FIG 2}. This yields a linear dependence where the slope can be related to the amplitude of the flux noise and the offset corresponds to the background dephasing level. In this case, instead of a simultaneous fit we apply a separate fit of $\Gamma_{\phi} = m D_{\Phi} + b$ to each qubit, and we find $A_{\Phi}^{1/2}$ values of $1.3 \pm 0.2$, $1.2 \pm 0.2$ and $1.4 \pm 0.2$~$\mu \Phi_0$ for the $\alpha = 7$, 4 and 1 qubits, respectively. These flux noise levels are all consistent with past studies of low-frequency flux noise in superconducting devices \citep{anton2012pure,yoshihara2006decoherence,bylander2011noise,bialczak2007flux,orgiazzi2016flux,stern2014flux}. 
%It is interesting to note that, although $A_{\Phi}^{1/2}$ calculated for these tunable qubits on sample A is the same within errors, the $\alpha = 1$ qubit appears to show the greatest $A_{\Phi}^{1/2}$. This is a consequence of fitting the Ramsey decays to a purely exponential envelope. From this value we put an upper bound on the absolute flux noise level in our samples.%, for a qubit strongly affected by $1/f$ flux noise. 

In Fig. \ref{FIG 4}a it can be seen that, for the tunable qubits on sample A, within the range $D_{\Phi} \lesssim 1\,{\rm GHz}/\Phi_0$, the measured dephasing rate is largely flux-independent within the experimental spread. To exploit this insensitivity, we designed the tunable transmon on sample B to have $D_{\Phi}$ no greater than $\sim 1\,{\rm GHz}/\Phi_0$ at any point within its tuning range, a condition satisfied by having $\alpha$ = 15. As a result, its sensitivity to $1/f$ flux noise appears to be suppressed below the level where background dephasing dominates. $\Gamma_{\phi}$ is essentially flat across the entire tuning range, as shown in Fig. \ref{FIG 3}, with a mean of 58 kHz and experimental scatter of $\sigma = $ 17~kHz. In comparison, this sample's fixed-frequency qubit exhibits $\Gamma_{\phi} = 72$~kHz. Figure \ref{FIG 4}b shows clearly that $T_2^*$ for the $\alpha=15$ qubit on sample B is independent of frequency over the whole tuning range. 
		
Although no significant flux dependence of the dephasing is detectable for sample B, we estimate from our earlier expression for $\Gamma_{\phi}$ that the observed scatter is consistent with $A_{\Phi}^{1/2}$ of 0.9~$\mu\Phi_0$. Recent progress in understanding the origins of $1/f$ flux noise in SQUIDs \cite{sendelbach2009complex} has facilitated up to a $5\times$ reduction in $A_{\Phi}$\cite{kumar2016origin}. Such reductions applied to the sample B qubit would reduce its maximum flux-noise-driven dephasing below 8~kHz. In a $\alpha = 7$ qubit tunable over more than 700~MHz, flux noise of such a level would cause dephasing no greater than 17~kHz. Alternatively, in a qubit of 150~MHz tunability, the flux noise seen in sample B would cause dephasing not exceeding 8~kHz, or only 4~kHz if the flux noise were reduced as in Ref. \cite{kumar2016origin}. We may contrast these values with the non-flux-noise-driven dephasing seen in state-of-the-art single-junction transmons used for multi-qubit gate operations: $\Gamma_{\phi} = $ 4 to 8~kHz on 2-qubit samples \cite{sheldon2016characterizing,Sheldon_Procedure_2016}, 10~kHz on 5-qubit samples \cite{ibmquantumexp} and 10 to 21~kHz on 7-qubit samples \cite{Takita2016Dem}.

In conclusion, we have shown that by reducing the flux-tunability of a transmon qubit, we can dramatically lower its sensitivity to $1/f$ flux noise. Using this understanding, we have fabricated a qubit in which the dephasing rate due to flux noise is suppressed below the level set by non-flux dependent sources. This device exhibits a flux-independent dephasing rate $\Gamma_{\phi} \sim 60\,{\rm kHz}$ over a tunable range in excess of $300\,{\rm MHz}$. This qubit design should be readily adaptable to existing architectures aimed at the realization of a logically-encoded qubit, in both frequency-tuned gates and all-microwave gates. As qubit architectures progress to more complex geometries, this work will enable the implementation of multi-qubit gates without frequency collisions impacting gate performance. This is a promising route to the creation of high-fidelity two-qubit gates for reaching fault tolerance, thus fundamentally improving the scalability of such systems for the creation of a universal quantum computer.

%This research was funded by the Office of the Director of National
%Intelligence (ODNI), Intelligence Advanced Research Projects Activity
%(IARPA), through the Army Research Office under Grant No. W911NF-16-0114 and Grant No. %W911NF-10-1-0324.
We acknowledge support from Intelligence Advanced Research Projects Activity (IARPA) under contract W911NF-16-0114.
The device fabrication was performed in part at the Cornell NanoScale Facility,
a member of the National Nanotechnology Coordinated Infrastructure
(NNIC) which is supported by the National Science Foundation under
Grant ECCS-1542081. We thank Y.-K.-K. Fung, J. Rohrs and J. R. Rozen for experimental contributions, and R. McDermott and T. Thorbeck for providing the SLUG amplifier and assistance with its setup. We also thank M. Brink, J.M. Gambetta, E. Magesan, R. McDermott, D.C. McKay and S. Rosenblatt for helpful discussions.
%
%We thank K. Fung, J. Rozen and J. Rohrs for assistance with apparatus and devices, and thank R. McDermott for helpful discussions.

%\bibliographystyle{apalike}
\bibliography{Main_bib}

\widetext
\clearpage
\begin{center}
\textbf{\large Supplementary Material for \textquoteleft Tunable Superconducting Qubits with Flux-Independent Coherencer"}
\end{center}

\setcounter{equation}{0}
\setcounter{figure}{0}
\setcounter{table}{0}
\setcounter{page}{1}
\makeatletter
\renewcommand{\theequation}{S\arabic{equation}}
\renewcommand{\thefigure}{S\arabic{figure}}
\renewcommand{\bibnumfmt}[1]{[S#1]}
\renewcommand{\citenumfont}[1]{S#1}

\section{Non-ideal fabrication in fixed frequency
qubits}

Lattices of coupled qubits are proposed to enable error-correction algorithms such as the `surface code' \cite{Gambetta2017Build_s,fowler2012surface_s}. Qubits are arranged into a square grid with alternate qubits serving either data or error-checking functions. Bus-couplers provide interaction among adjacent qubits, with up to four qubits attached to each bus. A seven qubit-lattice thereby comprises 12 qubit pairs and a seventeen-qubit lattice comprises 34 pairs. However, single junction transmon qubits are challenging to fabricate at precisely set frequencies. Among dozens of identically-fabricated qubits, the frequencies typically have a spread of $\sigma_f \sim 200$ MHz \cite{privcommsrosenblatt_s}. Such imprecision will inhibit functioning of qubit lattices. Considering a lattice of tansmon qubits of frequency $\sim 5$ GHz and anharmonicity $\delta/2\pi = -340$ MHz, and considering cross-resonance gate operations, we can estimate the number of undesired interactions among these pairs. Studies of the cross-resonance gate \citep{divincenzo2013quantum_s} indicate that these gates will be dominated by undesirable interactions if the frequency separation $|\Delta|$ between adjacent qubits is equal to zero, a degeneracy between $f_{01}$ of the qubits; equal to $-\delta/2\pi$, a degeneracy between $f_{01}$ of one qubit and $f_{12}$ of the next; or if $|\Delta| > -\delta/2\pi$ (weak interaction leading to very slow gate operation). In a simple Monte Carlo model, we assign to all points in the lattice a random qubit frequency from a gaussian distribution around 5 GHz, and count the number of degenerate or weak-interaction pairs, taking a range of $\pm (\delta/2\pi)/20$, or $\pm 17$ MHz around each degeneracy. The results appearing in Table \ref{table:MCModelCollisions} make it evident that the likelihood of frequency collisions increases as the lattice grows.

\begin{table}[h]
	\centering
	\begin{tabular}{c|c|c}
		Number &  & Mean Number \\
		of QBs & $\sigma_f$ & of Collisions \\ \hline
		7 & $\frac{1}{2}|\delta/2\pi|$ & 2.3  \\ 
		7 & $\frac{3}{4}|\delta/2\pi|$ & 3.6  \\ 
		17 & $\frac{1}{2}|\delta/2\pi|$ & 6.6  \\ 
		17 & $\frac{3}{4}|\delta/2\pi|$ & 10.6  
	\end{tabular} 
\caption{\label{table:MCModelCollisions} Frequency-collision modeling in lattices of transmon qubits employing cross-resonance gates. Predicted number of bad gate pairs (`frequency collisions') in two different lattice sizes. 7-qubit lattice has 12 pairs and 17-qubit lattice has 34 pairs. Mean of distribution is 5 GHz and two different distribution widths $\sigma_f$ are considered.}
\end{table}

\section{Device design and fabrication}

\begin{figure}[!b]
\includegraphics[width=1.0\columnwidth]{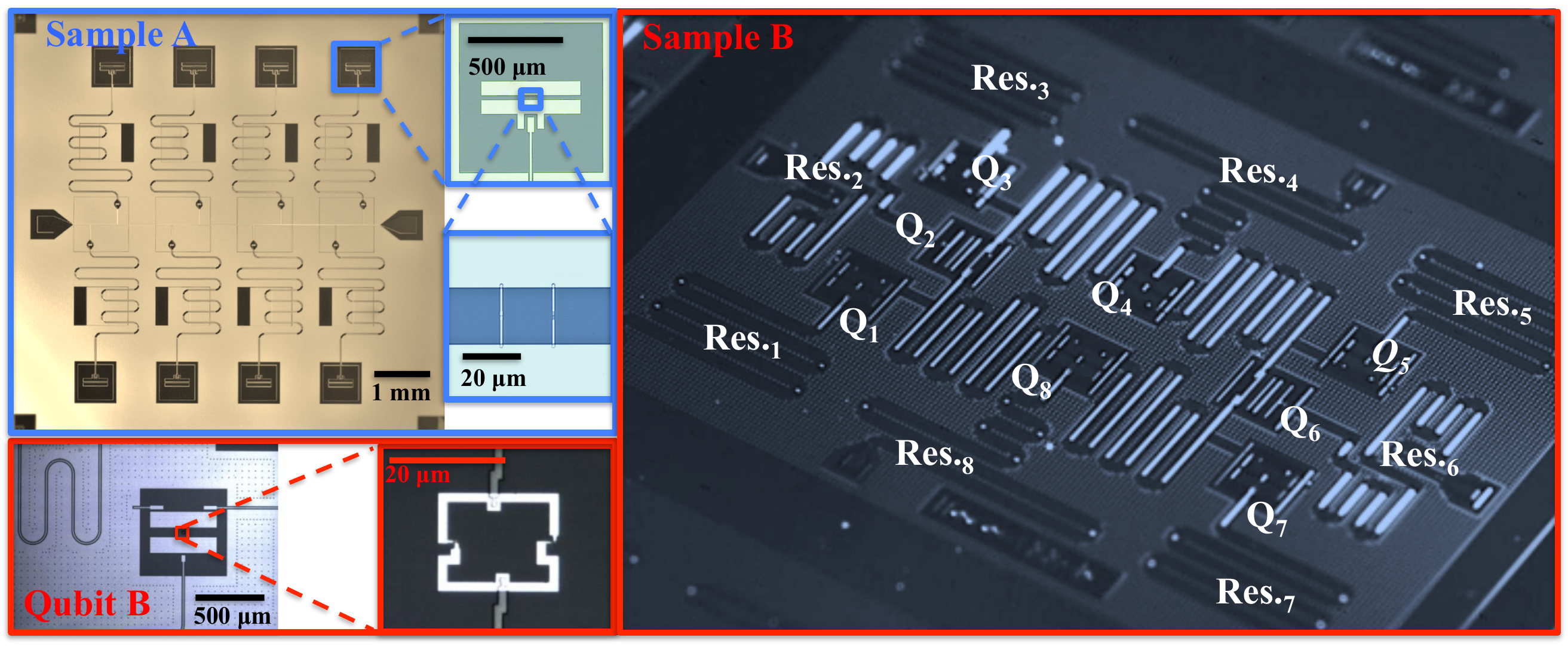}

\caption{(color online) Optical micrographs of samples including higher magnification images of qubits and SQUID loops. Sample B image is a chip of identical design to the ones used for measurements. In sample B image, labels indicate each qubit and its individual readout resonators, while unlabeled resonators are bus resonators. 
\label{fig:1}}
\end{figure}

The device for sample A, shown in Fig. \ref{fig:1}, has all
eight qubit/cavities capacitively coupled to a common feedline through
which individual qubit readout was achieved via a single microwave
drive and output line. Sample B, shown in Fig. \ref{fig:1}, employs a design where all qubits have separate drive and readout microwave lines. As in Ref. \cite{Takita2016Dem_s} and \cite{ibmquantumexp_s}, this sample is designed as a lattice of coupled qubits for use in multi-qubit gate operations, although no such operations are presented in this paper. Coplanar-waveguide buses, half-wave resonant at $\sim$6 GHz, span the space between the qubits. Each bus resonator couples together three adjacent qubits. As compared to Ref. \cite{Takita2016Dem_s}, here the lattice comprises eight qubits and four buses instead of the seven qubits and two buses found in Ref. \cite{Takita2016Dem_s}.

Both samples were fabricated using standard lithographic processing
to pattern the coplanar waveguides, ground plane, and qubit capacitors
from a sputtered Nb film on a Si substrate. In sample A the Nb films are 100 nm thick. In sample B they are 200 nm. The qubits were similar
in design to \citep{Sheldon_Procedure_2016_s,chow2014implementing_s,corcoles2015demonstration_s,Takita2016Dem_s} 
with large transmon capacitor pads bridged by electron-beam patterned Al traces used to
create Josephson junctions. Conventional shadow-evaporated double-angle Al-AlOx-Al was used to fabricate the junctions. Transmon capacitor pads in samples A and B have different size and separation, necessitating different SQUID loop geometries, as shown in Fig. \ref{fig:1}. The SQUID loops for qubits on sample A were created by bridging the transmon capacitor pads with
two separate $0.6-\mu{\rm m}$ wide Al traces and Josephson junctions, with the
asymmetry in the junctions fabricated by increasing the width of one
junction with respect to the other, while keeping the overlap fixed at $0.2\, \mu{\rm m}$. The sum of the large and small junction areas was designed to be constant, independent of $\alpha$.
Qubits on sample A had capacitor pads separated by $20\, \mu{\rm m}$ and
the Al electrodes separated such that the SQUID loop area was roughly
$400\, \mu{\rm m^{2}}$. In sample B, the Nb capacitor pads were separated by $70\, \mu{\rm m}$. The SQUID comprises a $\sim 20 \times 20\, \mu{\rm m}^2$ Al loop of 2 $\mu$m trace width, placed midway between the capacitor pads and joined to Nb leads extending from the pads. In sample B, the large and small junction differ in both width and overlap. In this sample, all SQUIDs of a given $\alpha$ were fabricated identically but SQUIDs of different $\alpha$ had different total junction area. 

\section{Measurement setup}

Measurements of sample A were completed in a dilution
refrigerator (DR) at Syracuse University (SU), while sample B was measured in a DR at the IBM TJ Watson Research Center. Both samples were wire-bonded into holders designed to suppress microwave
chip modes. Each sample was mounted to the mixing chamber of its respective DR and placed inside a cryoperm magnetic shield, thermally anchored at the mixing chamber. Both SU and IBM DRs had room-temperature $\mu$-metal shields. Measurements for both samples were performed using standard cQED readout techniques \citep{Reed_High_2010_s}.

For sample A, room-temperature microwave signals were supplied through attenuated coaxial
lines, thermalized at each stage of the DR and filtered using 10
GHz low pass filters (K\&L) thermalized at the mixing chamber. We used
a total of 70 dB of attenuation on the drive-lines: 20 dB at $4\, {\rm K}$, 20
dB at $0.7\, {\rm K}$ and 30 dB at the mixing chamber, with a base temperature of $30\,{\rm mK}$. Output measurement signals
from the sample pass through another 10 GHz low-pass filter, a microwave
switch, and two magnetically shielded cryogenic isolators, all thermally anchored
to the mixing chamber. In the case of sample A, the signal was amplified
by a low-noise HEMT at $4\, {\rm K}$, passing through a Nb/Nb superconducting
coaxial cable between the mixing chamber and $4\, {\rm K}$ stage. The signal was amplified further
at room temperature before being mixed down to 10 MHz and digitized. The eight resonators, coupled to each qubit on sample A, had measured frequencies that ranged from $6.975 - 7.136\, {\rm GHz}$, separated by $20 - 25\, {\rm MHz}$. $\kappa/{2\pi}$ linewidths for these resonators were on the order of a few hundreds of kHz.  

Figure \ref{fig:1} shows the layout of the sample B chip. The $\alpha = 15$ asymmetric-SQUID transmon reported in the paper was located at position $Q_7$. It was read out through a coplanar waveguide resonator of frequency 6.559 GHz and linewidth $\sim$ 300 kHz, and was found to have $f_{01}^{max} = 5.387$ GHz. The fixed-frequency transmon (5.346 GHz) at position $Q_2$ was read out through a 6.418 GHz resonator having linewidth $\sim$ 300 kHz. Sample B qubits were measured via signal wiring similar to that presented in Refs. \cite{Takita2016Dem_s,chow2014implementing_s,corcoles2015demonstration_s,sheldon2016characterizing_s}. Drive wiring included 10 dB of attenuation at 50 K, 10 dB at 4K, 6 dB at 0.7 K, 10 dB at 100 mK, and at the mixing-chamber plate 30 dB of attenuation plus a homemade `Eccosorb' low-pass filter. Drive signals entered a microwave circulator at the mixing plate. On one set of signal wiring, the 2nd port of the circulator passed directly to qubit $Q_7$. In another set of signal wiring, the second port of the circulator passed to several different qubits via a microwave switch. Signals reflected from the device passed back through the circulator to output and amplifier circuitry. Output circuitry comprised a low-pass Cu powder filter, followed by two cryogenic isolators in series, followed by an additional low-pass filter, followed by superconducting NbTi coaxial cable, followed by a low-noise HEMT amplifier at 4K and an additional low-noise amplifier at room temperature. Low-pass filters were intended to block signals above $\sim$ 10 GHz. In the case of $Q_7$, additional amplification was afforded by a SLUG amplifier \cite{HoverAPL2014_104_152601_s} mounted at the mixing stage, biased via two bias-tee networks and isolated from the sample by an additional cryogenic isolator. Output signals were mixed down to 5 MHz before being digitized and averaged. Mixing-plate thermometer indicated a temperature of $\sim$ 15 to 20 mK during measurements. 

Magnetic flux was supplied to sample A via a 6-mm inner diameter superconducting
wire coil placed $2\,{\rm mm}$ above the sample. A Stanford SRS SIM928 dc voltage source with a room-temperature $2\,{\rm k}\Omega$ resistor in series supplied the bias current to the coil. The flux bias current passed through brass coaxial lines that were thermally anchored
at each stage of the DR, with a $80~{\rm MHz}$ $\pi$-filter at 4K and a copper powder filter on the mixing chamber. In sample B, a similar wire-wound superconducting coil was mounted about 3 mm above the qubit chip and likewise driven from a SIM928 voltage source through a room-temperature $5\,{\rm k}\Omega$ bias resistor. DC pair wiring (Cu above 4K within the fridge, NbTi below) was used to drive the coil. The coil had a self-inductance of 3.9 mH and mutual inductance to the SQUID loop of $\sim$ 1 pH. The flux coil applied a dc flux through
all qubits with the flux level being set just prior to qubit measurement
and maintained at a constant level throughout the measurement. For each qubit, we measured $f_{01}$ as a function of coil current and fit this against Eq. (1) of our paper to enable scaling of $\Phi_0$ and subtract any offset flux, as well as to determine $f_{01}^{max}$ and asymmetry $d$. We treat the sign of flux as arbitrary. 

\section{Qubit Coherence}

Coherence data for both samples was collected using an automated measurement algorithm. After applying a prescribed fixed flux, the system determined the qubit frequency from Ramsey fringe fitting, optimized $\pi$ and $\pi/2$ pulses at this frequency, and measured coherence. $T_{2}^{*}$ measurements were completed at a frequency detuned from the qubit frequency, with the level of detuning optimized to provide a reasonable number of fringes for fitting. All raw coherence data was visually checked to confirm that a good quality measurement was achieved. If the automated tuning routine failed to find the frequency or properly scale the $\pi$ and $\pi/2$ pulses, this point was omitted from the dataset.
For sample A, three $T_1$ measurements were made at each flux point followed by three $T_2^*$ measurements. At each flux point, the reported $T_1$ and $T_2^*$ values and error bars comprise the mean and standard deviation of the three measurements. The corresponding $\Gamma_{\phi}$ value is found from these mean values and its error bar is found by propagating the errors in $T_1$ and $T_2^*$ through via partial derivative and combining these in a quadrature sum. For sample B, at each flux point first $T_1$ was measured, then $T_2^*$, three times in succession. For this device the reported $T_1$ and $T_2^*$ values comprise the mean of the three measurements and the error bars are their standard deviation. Here the reported dephasing rate $\Gamma_{\phi}$ comprises the mean of the three values of $\Gamma_{\phi}=1/T_{2}^{*}-1/2T_{1}$ found from the three $T_1$, $T_2^*$ pairs, and the error bar is the standard deviation. 

\begin{figure}
\includegraphics[width=0.75\columnwidth]{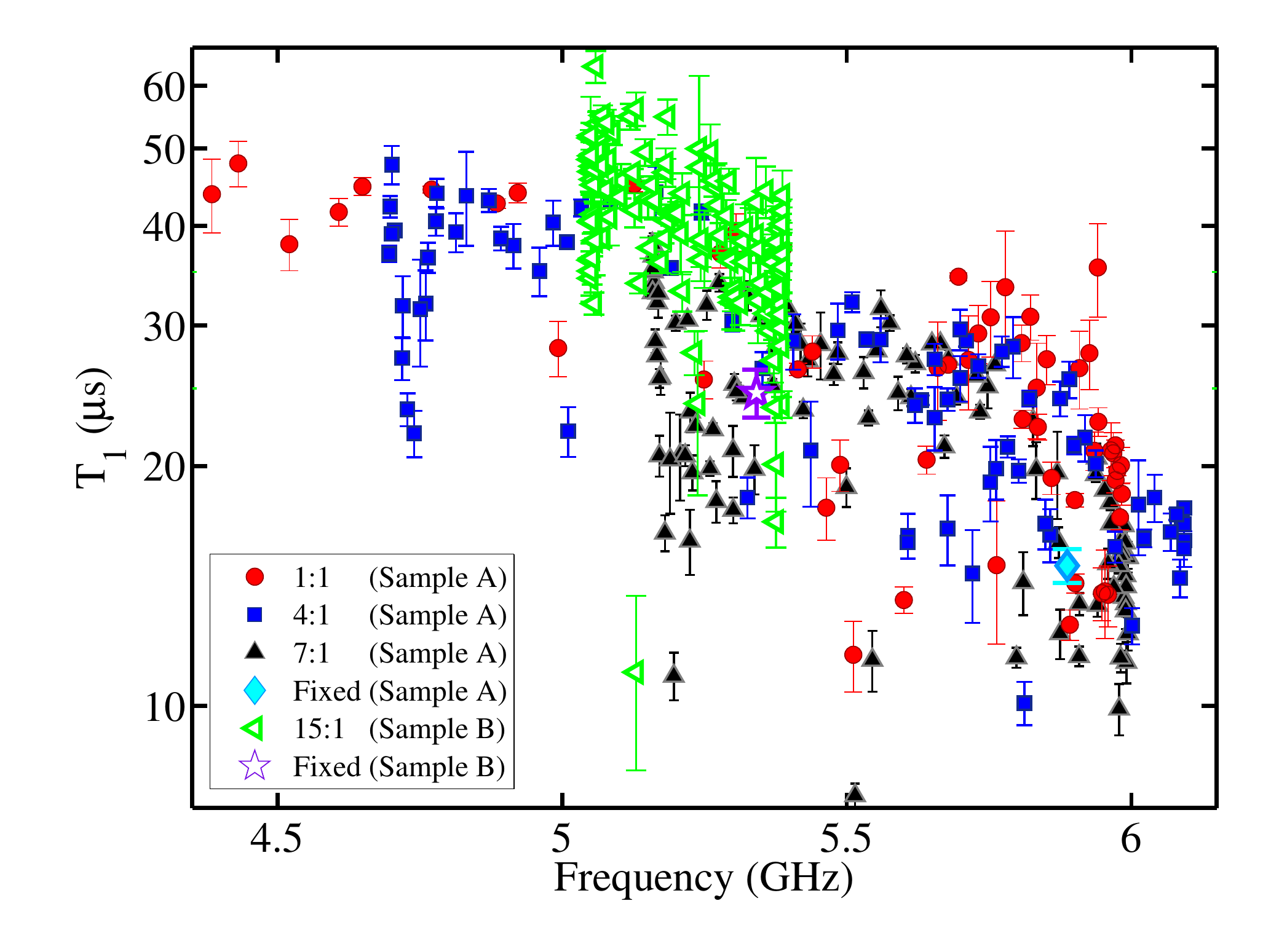}

\caption{$T_{1}$ vs. frequency measured for all qubits discussed in the main paper. Single points included for $T_{1}$ values measured for the fixed-frequency qubits. 
\label{fig:2}}

\end{figure}

Figure \ref{fig:2} shows $T_{1}$ plotted versus qubit frequency, measured for the qubits discussed in our paper. We observe a trend
of increasing  $T_{1}$ with decreasing qubit frequency. In sample A, each qubit's quality factor $\omega T_{1}$ is roughly constant, consistent with dielectric loss and a frequency-independent loss tangent, as observed in other tunable superconducting qubits \citep{barends2013coherent_s}. On sample B, $T_{1}$ decreases by about 10 $\mu$s from the low to high end of the frequency range, consistent with Purcell loss to the readout resonator. In addition, fine structure is occasionally observed in Fig.
\ref{fig:2} where $T_{1}$ drops sharply at specific frequencies. These localized features in the $T_{1}$ frequency dependence are observed
for all tunable qubits that we have measured. These features, similar to those observed by \citep{barends2013coherent_s}, are attributed to frequencies where a qubit transition is resonant with a two-level system defect on or near the qubit. Additionally, on sample B, at a few frequency points inter-qubit coupling affects relaxation. Where the $Q_7$ qubit is nearly degenerate to $Q_6$ (at $\sim$5.33 GHz) and to $Q_8$ (at $\sim$5.22 GHz), coupling via the adjacent buses produces an avoided crossing in the energy spectrum. This effect is barely noticeable in both the frequency curve of Fig. 2 of our paper as well as the relaxation data in Fig. \ref{eq:2} here.

\begin{figure}
\includegraphics[width=0.75\columnwidth]{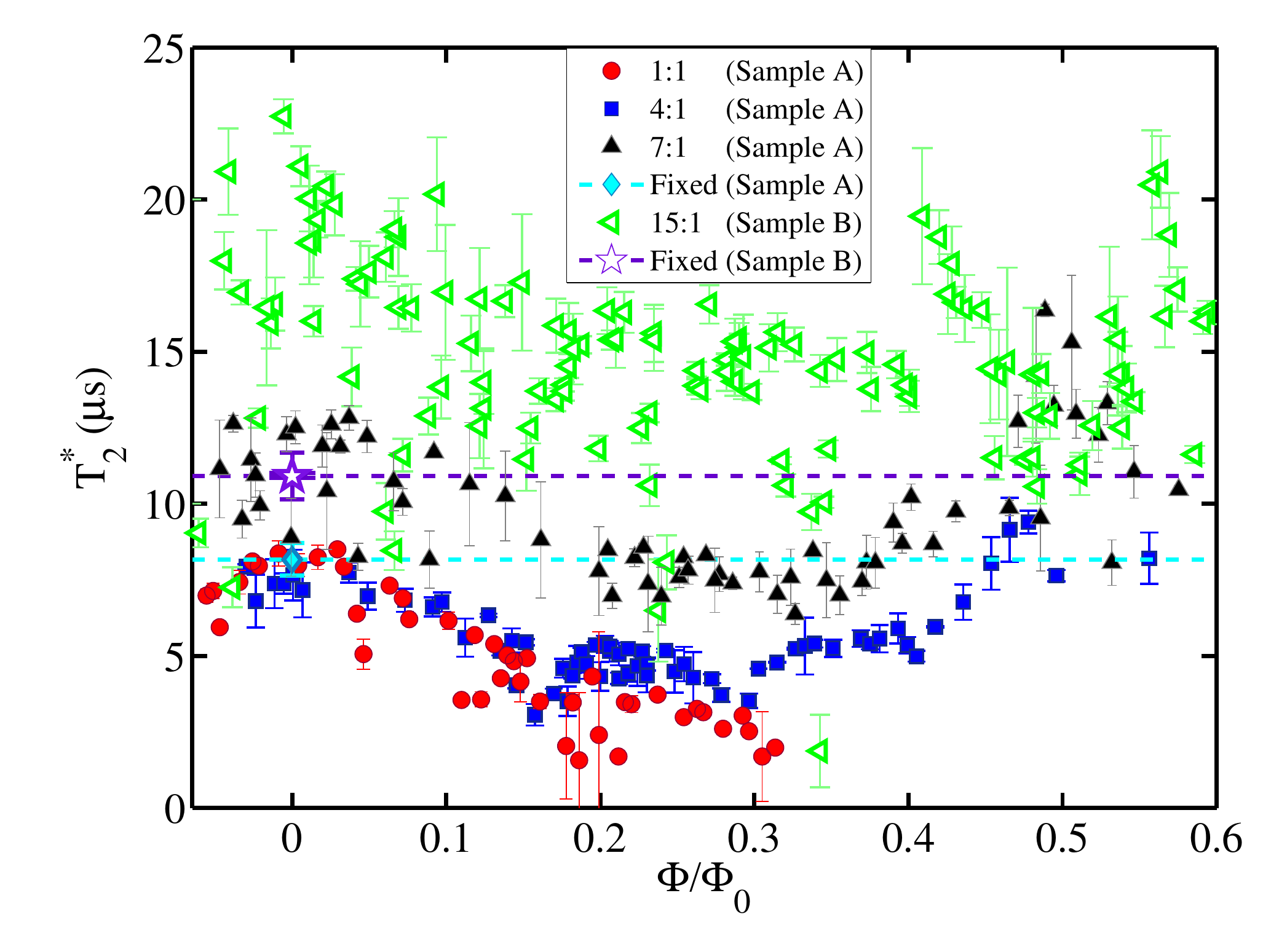}

\caption{$T_{2}^{*}$ vs. flux measured for the qubits discussed in the main paper. $T_{2}^{*}$ measured for the fixed-frequency qubits on both samples is included with dashed lines to help guide the eye.
\label{fig:3}}

\end{figure}

Figure \ref{fig:3} shows $T_{2}^{*}$ plotted versus flux, measured
for the qubits discussed in our paper.
For the tunable qubits on
sample A, $T_{2}^{*}$ is greatest at the qubit sweet-spots and decreases
away from these sweet spots as $D_{\Phi}$ increases.
In the $\alpha = 15$ tunable qubit on sample B, $T_{2}^{*}$ is nearly constant over the measured half flux quantum range. The small frequency dependence observed in $T_{2}^{*}$ in sample B is consistent with the observed variation of $T_{1}$ with frequency, leading to the frequency-independent dephasing rate observed for this qubit
in Fig. 3 of our paper.

\section{Relaxation Due to coupling to Flux Bias Line}

\begin{figure}

\includegraphics[width=0.75\columnwidth]{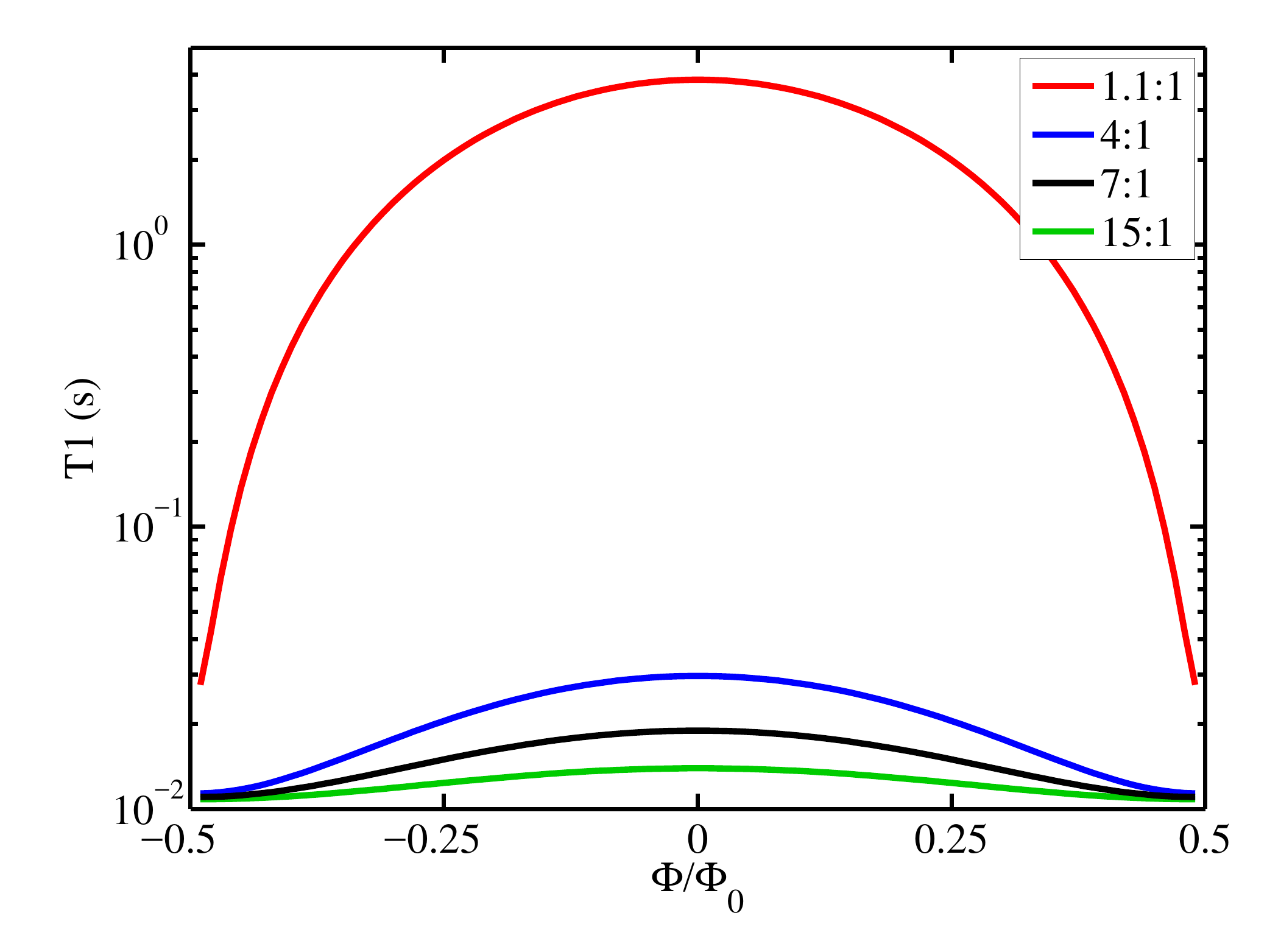}

\caption{Dependence of $T_{1}$  with flux for asymmetric transmons, calculated for the asymmetries discussed in the main paper, due to coupling to an external flux bias following the analysis of Koch et al \citep{koch2007charge_s}. Though in the main paper our symmetric qubit was an $\alpha = 1$, in this calculation we used $\alpha = 1.1$ so that $T_{1}$ did not diverge at $\Phi = 0$. 
\label{fig:4}}

\end{figure}

While using two Josephson junctions to form a dc SQUID for the inductive element of a transmon allows its frequency to be tuned via magnetic flux, this opens up an additional channel for
energy relaxation via emission into the dissipative environment across the bias coil that is coupled to the qubit through a mutual inductance. This was first discussed by Koch et al \citep{koch2007charge_s}. regarding
a near symmetrical split-junction transmon. We apply the same analysis
here to study the effect of increasing junction asymmetry on the qubit $T_{1}$ through this loss mechanism. For an asymmetric transmon, Koch et al. show in Eq. {(}2.17{)} of Ref. \citep{koch2007charge_s} that the Josephson portion of the qubit Hamiltonian can be written in terms of a single phase variable with a shifted minimum that depends
upon the qubit's asymmetry and the applied flux bias. 
By linearizing this Hamiltonian about the static flux bias point for small noise amplitudes, Koch et al. compute the relaxation rate for a particular current noise power from the bias impedance coupled to the SQUID loop through a mutual inductance $M$.
%By assuming that the flux bias drive consists of some external flux plus a small noise
%term, this Hamiltonian can be expanded to include this noise term.
%Treating the noise spectrum perturbatively, qubit relaxation rate
%can be related to the noise power spectrum using Eq 4.13 in kock et al.
We followed this same analysis for our qubit parameters, assuming harmonic oscillator wavefunctions for the qubit ground and excited state, and obtained the dependence of $T_{1}$ due to this mechanism as a function of bias flux.
%By assuming a hormonic oscillator wavefunction for the qubit, we
%can gain an order of magnitude estimate of the relaxation due to the
%flux line mutual inductance using Eq 4.13. 
Using our typical device
parameters ($E_{J} = 20\,{\rm GHz}$, $E_{c} = 350\,{\rm MHz}$, $M = 2\,{\rm pH}$, $R =
50~\Omega$) we obtain the intrinsic loss for the asymmetries
discussed in our paper, shown in Fig. \ref{fig:4}. This analysis
agrees with the results described in Ref. {[}4{]}. For a 10\% junction asymmetry, this contribution results in a $T_{1}$ that varies between $25\,{\rm ms}$ and a few seconds. As the junction asymmetry is increased,
the minimum $T_{1}$ value, obtained at odd half-integer multiples of $\Phi_{0}$, decreases slightly. However, even for our $\alpha = 15$ qubit, the calculated value of $T_{1}$ due to this mechanism never falls below $10\,{\rm ms}$. Therefore, although increasing
junction asymmetry does place an upper bound on $T_{1}$ of an asymmetric transmon, this level is two orders of magnitude larger than the measured $T_{1}$ in current state-of-the-art superconducting qubits due to other mechanisms.

Also in Ref. [4], Koch et al. described a second loss channel for a transmon related to coupling to the flux-bias line. In this case, the relaxation occurs due to the oscillatory current through the inductive element of the qubit -- independent of the presence of a SQUID loop -- coupling to the flux-bias line, described by an effective mutual inductance $M'$. This mutual vanishes when the Josephson element of the qubit and the bias line are arranged symmetrically. With a moderate coupling asymmetry for an on-chip bias line, Koch et al. estimate that the $T_{1}$ corresponding to this loss mechanism would be of the order of 70 ms. Because this mechanism does not directly involve the presence or absence of a SQUID loop for the inductive element, the asymmetry between junctions that we employ in our asymmetric transmons will not play any role here and this particular limit on $T_{1}$ should be no different from that for a conventional transmon. An additional potential relaxation channel may arise due to capacitive coupling to the flux-bias line, as discussed in Ref. \cite{JohnsonDisserationYale2011_s}. However, this is expected to be negligible where a bobbin coil is used as in our experiments.
%Also in Ref. {[}4{]}, Koch el al. also discussed a second decay channel wherein the entire
%transmon circuit couples to the flux bias line through a separate mutual
%inductance $M\textquoteright$. This mutual vanishes for device
%geometries where the SQUID loop and bias line is exactly centered
%in the middle of the transmission line resonator. Therefore, this
%mutual is only important when realizing a flux bias line with an on-chip
%coplanar waveguide, displaced with respect to the qubit to maximize
%coupling to the SQUID loop. They show that, for the same realistic
%parameters used to calculate relaxation due to the primary mutual
%$M$, this mutual $M\textquoteright$ would create relaxation times
%on the order of 10s of ms. Asymmetry does not impact this relaxation
%time so we conclude that this would not impact a scalable solution
%to quantum computing that utilizes on-chip flux lines and weakly tunable
%qubits.

\section{Ramsey Decay Fitting}

As described in the main paper, our analysis of qubit dephasing rates used a purely exponential fit
to all of the measured Ramsey decays. Here we discuss why this fitting approach is appropriate for all asymmetric qubits and a large portion of
the coherence data measured for the symmetric qubit.

\begin{figure}

\includegraphics[width=1\columnwidth]{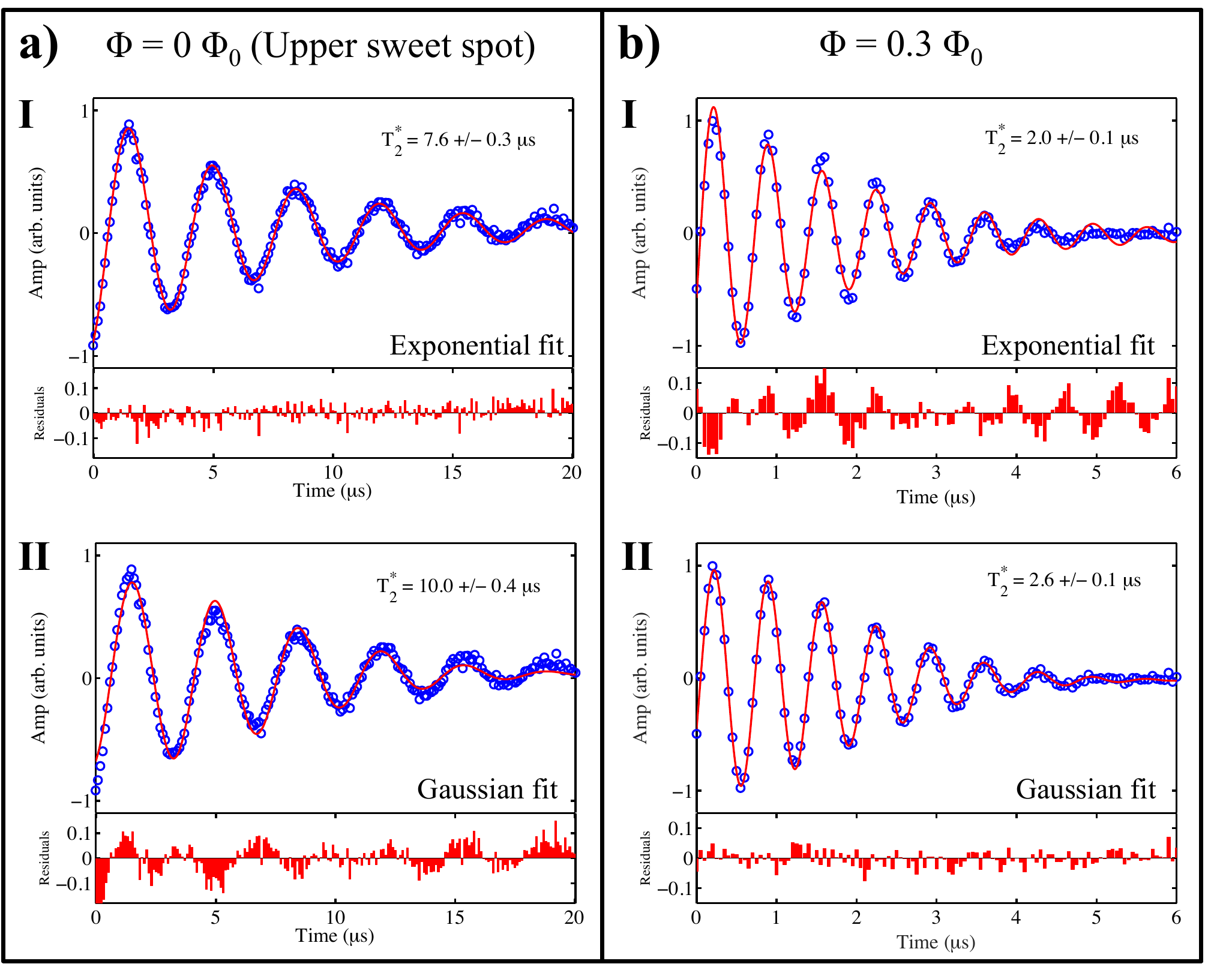}
\caption{Ramsey decay envolopes measured for the $\alpha = 1$ qubit at a) the sweet-spot $\Phi=0$ and b) $\Phi=0.3 \Phi_{0}$ where $D_{\Phi}$ was the largest value measured for this qubit. At each flux point, the Ramsey decay envelopes are fit with both a purely exponential {(}I{)} and Gaussian {(}II{)} fit form. Functions fitted to the measured data {(}blue open circles{)} plotted as solid red lines.
\label{fig:5}}

\end{figure}

Of all the qubits measured in this study, the symmetric $\alpha = 1$ qubit
was most impacted by flux noise away from the qubit sweet spot because of its large energy-band gradient. Therefore,
to illustrate the impact that flux noise has upon the Ramsey decay
envelope we will consider the Ramsey measurements for this qubit on
and off the sweet spot. Example measurements are shown at flux values of 0 and 0.3 ~$\Phi_0$ in Fig. \ref{fig:5}a and b, respectively. At each flux point,
we fit the Ramsey decay with both a purely exponential (Fig. \ref{fig:5}a
I) and purely Gaussian form (Fig \ref{fig:5}a II), the residuals
of each fit are included to compare the quality of fit in each case.
As has been discussed in the main paper, at the upper sweet-spot, where
$D_{\Phi} = 0$, non-flux dependent background-dephasing
should dominate and the Ramsey decay should be more readily fit using
an exponential. Figure \ref{fig:5}a shows that this is indeed the
case: the purely exponential fit provides a more precise fit to the
Ramsey decay, with the residuals to this fit being smaller over the entire range compared to those corresponding to the Gaussian fit. The Ramsey decay
shown in Fig \ref{fig:5}b was measured at a point where $D_{\Phi}$
was the maximum measured for the $\alpha = 1$ qubit. Here, it is clear that
a purely Gaussian form results in a better fit with smaller residuals than an exponential envelope. This indicates that, at this flux point, the $\alpha = 1$ qubit is heavily impacted by low-frequency flux noise, as a purely 1/f dephasing source would result in a Gaussian envolope for the decay \citep{ithier2005decoherence_s}. Although a purely Gaussian fit form is useful
for illustrating the impact that flux noise has upon the Ramsey decay
form, it is not an optimal quantitative approach for investigating dephasing in these qubits. This is because tunable transmons dephase not only due to flux noise with a roughly $1/f$ power spectrum, but also due to other noise sources with different non-$1/f$ power spectra \citep{sears2012photon_s,schuster2005ac_s,gambetta2006qubit_s}. These other noise sources generally result in an exponential dephasing envelope. Also, dephasing has an intrinsic loss component that is always exponential in nature. Therefore, to accurately fit decay due to dephasing in these qubits, we must account for these exponential decay envelopes in any fitting approach that is not purely exponential.

To account for the $T_{1}$ contribution to the Ramsey decay envelope in our non-exponential fitting, we take the average $T_{1}$ measured at each flux point
and separate this from $T_{2}^{*}$ in the Ramsey fit function using
$1/T_{2}^{*} = 1/T_{\phi} + 1/2T_{1}$. Therefore, instead
of fitting a $T_{2}^{*}$ time, we fit $T_{\phi} $
directly. To fit the Ramsey using a Gaussian fit form, we square the
dephasing exponent within the fitting function {[}Eq. {(}\ref{eq:1}{)}{]}. We can go one step
further by not forcing an explicit fit form to the dephasing exponent, but instead adding another fit parameter $\gamma$ {[}Eq. {(}\ref{eq:2}{)}{]}, which would be 1 for a pure exponential and 2 for a pure Gaussian. Although a fit that is not explicitly exponential or Gaussian is not motivated directly by a particular theoretical model, by fitting Ramsey decays with this free exponent
$\gamma$, we gain insight into the transition from flux-noise dominated dephasing at large $D_{\Phi}$ to background dephasing near the sweet-spots. The two separate fit forms described above are given by the following decay functions:

\begin{equation}
f_{Ramsey}(t)=A+B\{\cos{(\omega t+\delta)}\exp{(-\Gamma_{1}t/2)}\exp{[-(\Gamma_{\phi}t)^2]}\}\label{eq:1},
\end{equation}
\begin{equation}
f_{Ramsey}(t)=A+B\{\cos{(\omega t+\delta)}\exp{(-\Gamma_{1}t/2)}\exp{[-(\Gamma_{\phi}t)^\gamma}]\}\label{eq:2},
\end{equation}
where A and B are magnitude and offset constants to adjust the arbitrary measured signal, $\omega$ is the detuning from the qubit frequency with a phase offset $\delta$, $\Gamma_{1}$ is the intrinsic loss rate {(}$1/T_{1}${)} and $\Gamma_{\phi}$ is the dephasing rate. Here, A, B, $\omega$, $\delta$, $\Gamma_{\phi}$, and $\gamma$ are fit parameters. All other components are fixed with values determined using the methods discussed above.

\begin{figure}

\includegraphics[width=0.75\columnwidth]{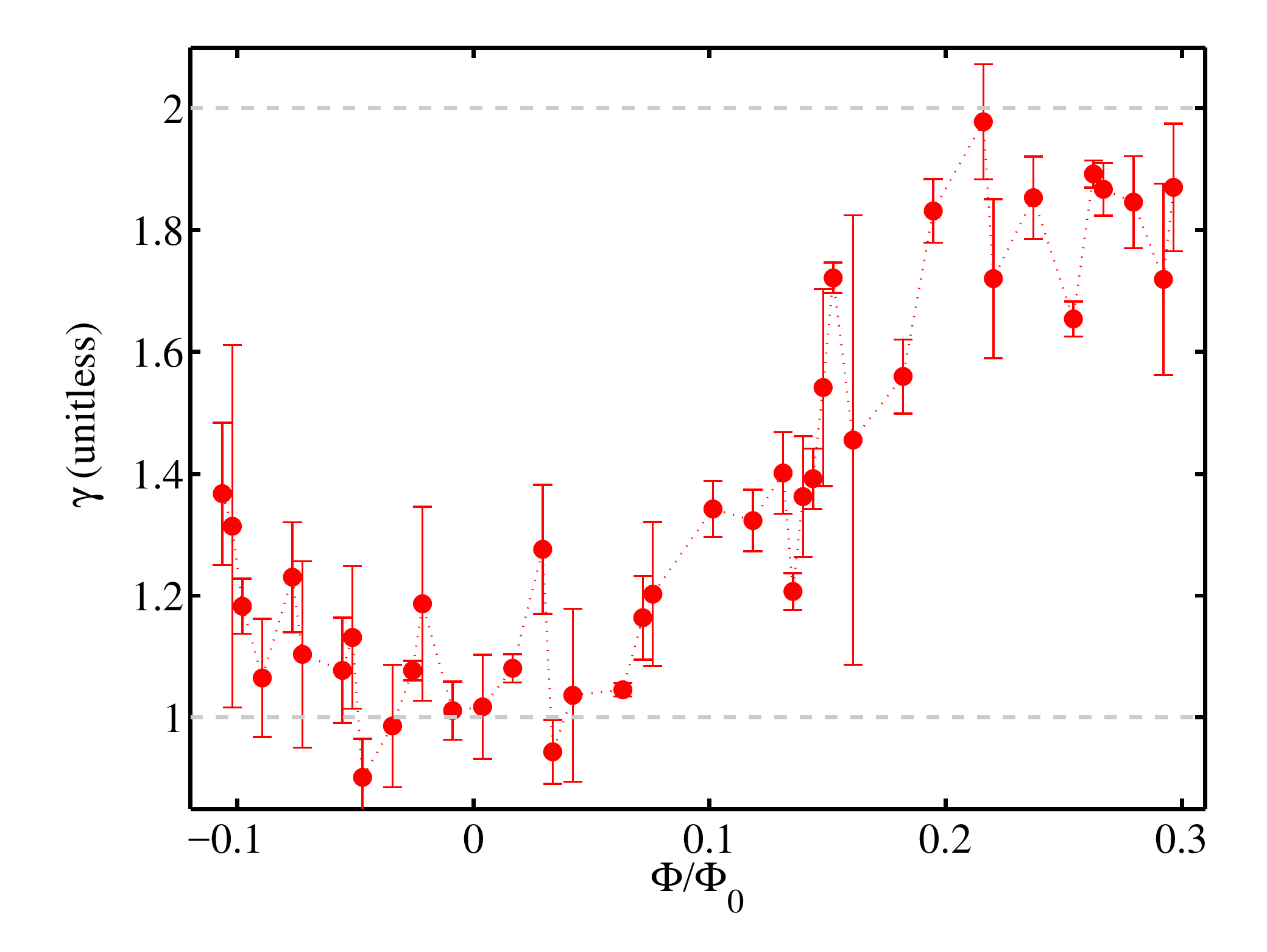}
\caption{$\gamma$ vs flux extracted from fits to the Ramsey measurements on the $\alpha = 1$ qubit using Eq. \ref{eq:2}.
\label{fig:6}}

\end{figure}

This behavior is illustrated in Fig. \ref{fig:6}, where we plot $\gamma$
vs. flux extracted from fits to the Ramsey measurements on the $\alpha = 1$
qubit using Eq. {(}\ref{eq:1}{)}. In the flux region between +/- 0.1 $\Phi_{0}$, $\gamma \approx 1$,
indicating that the dephasing envelope is primarily exponential, and thus the dominant dephasing noise affecting the qubits here does not have a $1/f$ spectrum. At flux bias points further away from the sweet-spot, $\gamma$ shifts towards 2 as $D_{\Phi}$ increases and appears
to level off close to this value at flux biases above $\sim 0.2\,\Phi_{0}$. Thus, in this bias regime, the dephasing envelope is primarily Gaussian and the dephasing noise influencing the qubits is predominantly low-frequency in nature with a $1/f$-like spectrum \citep{ithier2005decoherence_s,yoshihara2006decoherence_s}. 

\begin{figure}
\includegraphics[width=0.75\columnwidth]{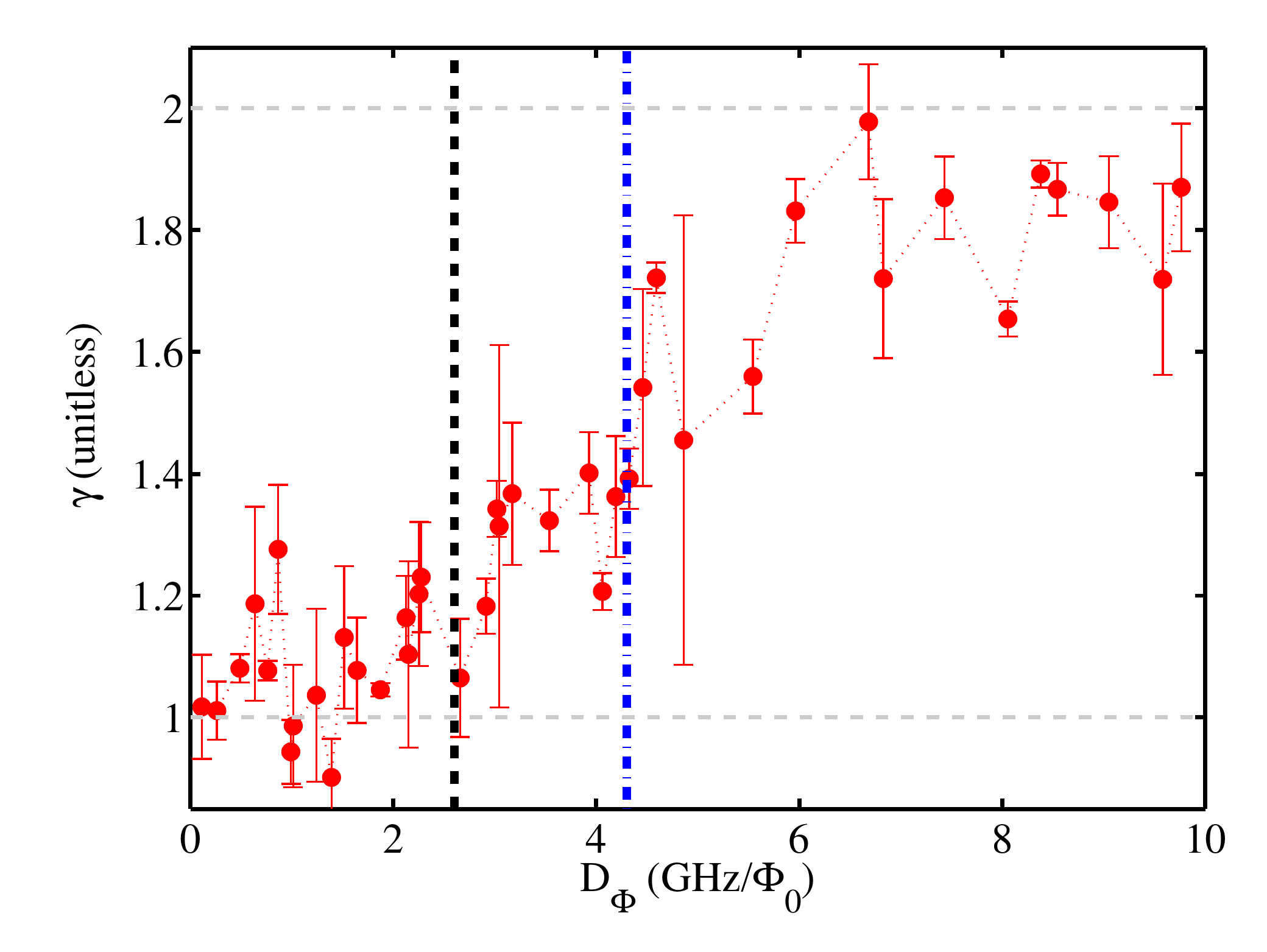}

\caption{$\gamma$ vs $D_{\Phi}$ extracted from fits to the Ramsey measurements on the $\alpha = 1$ qubit using Eq. \ref{eq:2}. Dashed lines included to indicate the maximum $D_{\Phi}$ reached by the $\alpha = 7$ {(}black dashed line{)} and $\alpha = 4$ {(}blue dot-dashed line{)} qubits measured on sample A.
\label{fig:7}}

\end{figure}

We can also vizualize this variable-exponent fit by plotting $\gamma$ vs. $D_{\Phi}$ rather than $\Phi$, again, for the $\alpha = 1$ qubit {(}Fig. \ref{fig:7}{)}.
%This study revealed how a qubits dephasing rate is proportional to $D_{phi}$. This means that as the rate of change of qubit frequency increases,
%its sensitivity to $1/f$ flux noise increases. To illustrate this,
%we plot $\gamma$ vs. $D_{\Phi}$.
%for the $\alpha = 1$ qubit in Fig. \ref{fig:7}. 
In this plot, $\gamma$ approaches 2 for $D_{\Phi}$ values around $6~\rm GHz/\Phi_{0}$. We have also included vertical dashed lines on Fig. \ref{fig:7} indicating the maximum $D_{\Phi}$ values reached by the less tunable $\alpha = 4$ and 7 qubits on sample A. Below these $D_{\Phi}$ levels, $\gamma$ is close to 1 implying that the decay envelope is nearly exponential, and thus justifying our use of an exponential decay for fitting the asymmetrical qubits in the main paper.

%This $\gamma$-exponent fit {[}of form Eq. {[}?{]}{]} is useful for illustrating how the Ramsey
%decay envelope evolves from a more exponential to Gaussian form as
%the different non-flux and flux dependent dephasing process become
%more or less dominant. 
As yet another approach to fitting the Ramsey decay envelopes, we can employ a function that separates the exponential from background-dephasing from the Gaussian form due to dephasing from noise with a low-frequency tail.
%As mentioned earlier, a decay function that
%is not explicitly exponential or Gaussian has no physical
%meaning. Using the understanding that non-flux dependent background
%dephasing processes dominate at the qubits upper sweet spot, we define
%a fourth Ramsey decay fit form that should provide a firmer physical
%model of decay due to dephasing. 
For this fit, along with separating
out the $T_{1}$ contribution to the Ramsey decay envelope, we also determine the
non-flux dependent background-dephasing rate at the sweet-spot,
then use this rate as a fixed parameter in the fitting of our Ramsey measurements
at any given flux point. We now have a composite Ramsey fit form that
has three components: a $T_{1}$ contribution and background dephasing component that are purely exponential and fixed by the fitting of separate measurements, plus a Gaussian component to capture the dephasing due to noise with a $1/f$ spectrum. This leads to a composite fitting function of the form:

\begin{equation}
f_{Ramsey}(t)=A+B\{\cos{(\omega t+\delta)}\exp{(-\Gamma_{1}t/2)}\exp{(-\Gamma_{\phi ,bkg}t)}\exp{[-(\Gamma_{\phi}t)^2}]\}\label{eq:3},
\end{equation}
where A and B are magnitude and offset constants to adjust the arbitrary measured signal, $\omega$ is the detuning from the qubit frequency with a phase offset $\delta$, $\Gamma_{1}$ is the intrinsic loss rate {(}$1/T_{1}${)}, $\Gamma_{\phi ,bkg}$ is the background dephasing rate measured at $D_{\Phi}=0$ and $\Gamma_{\phi}$ is the fitted dephasing rate. Here, A, B, $\omega$, $\delta$, and $\Gamma_{\phi}$ are fit parameters. All other components are fixed with values determined using methods discussed above. Though this fit form well separates the different components to dephasing decay, it has one key deficiency: it assumes that the background dephasing rate is frequency independent, which is not necessarily justified, as the background dephasing mechanism may also vary with frequency. To calculate the total dephasing rate using this fit form, we add the constant background dephasing to the fitted $\Gamma_{\phi}$. 

\begin{figure}

\includegraphics[width=0.75\columnwidth]{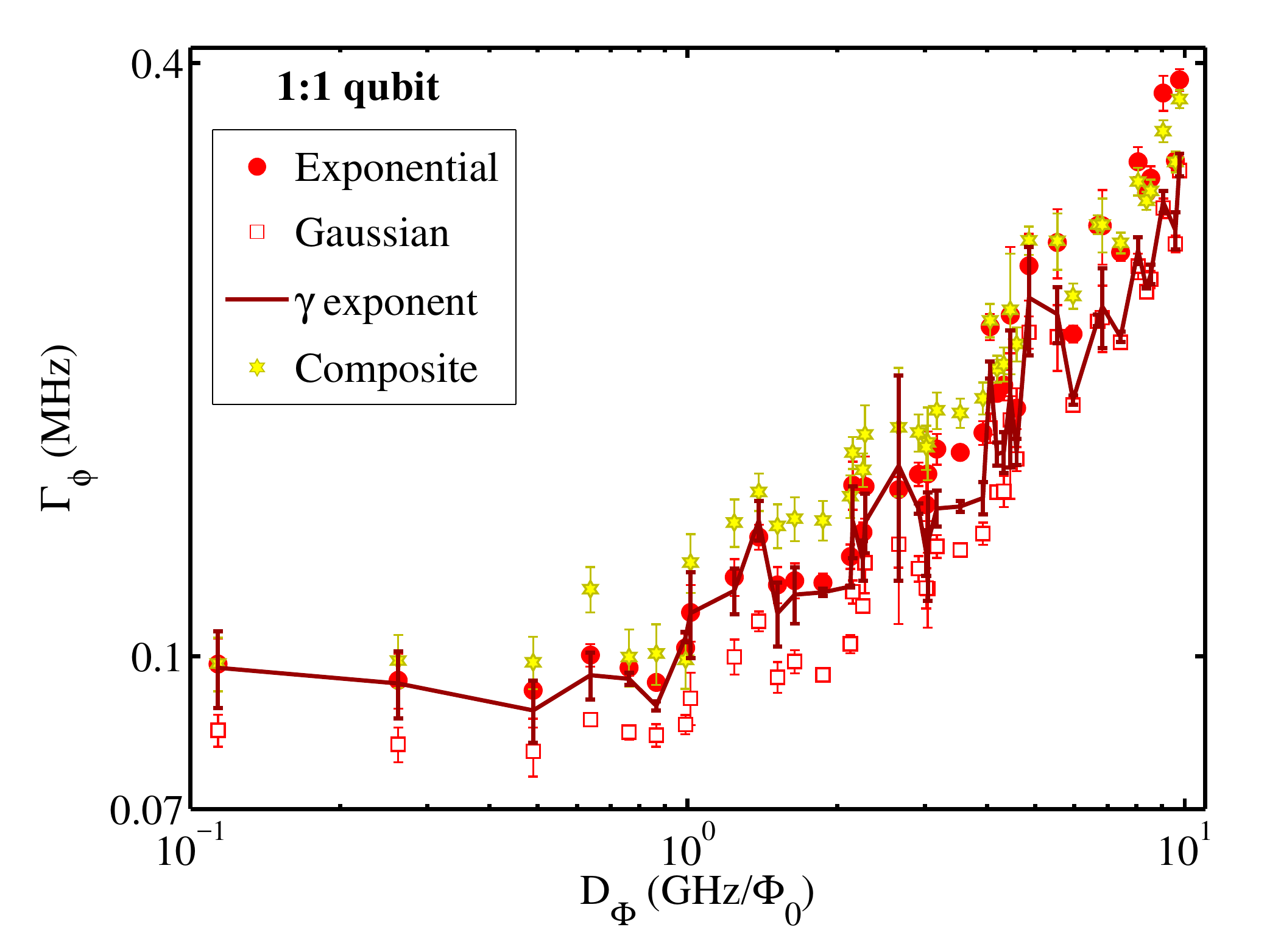}
\caption{$\Gamma_{\phi}$ vs. $D_{\Phi}$ calculated for the $\alpha = 1$ qubit using the exponential,
Gaussian {[}Eq. {(}\ref{eq:1}{)}{]}, $\gamma$-exponent {[}Eq. {(}\ref{eq:2}{)}{]}, and composite {[}Eq. {(}\ref{eq:3}{)}{]}fitting forms.
\label{fig:8}}
\end{figure}

To understand how the explicit fitting form impacts the dephasing rate, in Fig. \ref{fig:8} we plot $\Gamma_{\phi}$ vs. $D_{\Phi}$ calculated for
the $\alpha = 1$ qubit using the four different fitting forms: exponential, Gaussian {[}Eq. {(}\ref{eq:1}{)}{]}, $\gamma$-exponent {[}Eq. {(}\ref{eq:2}{)}{]}, and composite {[}Eq. {(}\ref{eq:3}{)}{]}. We first note that any differences in the rate of dephasing calculated at each point using the various fit methods
are subtle and the fits are reasonably consistent with one another within the fit error bars and scatter. We do observe, though, that a purely exponential fit results in
a dephasing rate that is slightly higher than the values from the Guassian fits for all flux points, resulting in the largest slope and thus the highest effective flux-noise level. Therefore, we conclude that forcing a purely exponential fit to the Ramsey decay envelopes measured for qubits that are strongly influenced
by $1/f$ flux noise simply puts an upper bound on the absolute flux
noise strength. The $\gamma$-exponent fitting approach provides a dephasing
rate that agrees well with that extracted from the exponential fit
form at low $D_{\Phi}$ values where background-dephasing
processes dominate. However, at higher $D_{\Phi}$ values
where the qubit is heavily impacted by $1/f$ flux noise, the $\gamma$-exponent
fit provides better agreement with the Gaussian-fitted dephasing rate.

The composite fit is rigidly fixed in the $\Gamma_{\phi}$ axis by the value chosen
to match the background dephasing rate, in this case chosen to match
the rate observed at the lowest $D_{\Phi}$ for the pure
exponential fit. For this reason, direct comparisons between this
fit and the others at individual flux points is more difficult. Despite all of these potential issues, the slope of $\Gamma_{\phi}$ vs. $D_{\Phi}$ is independent of the chosen background-dephasing
rate. Therefore, this composite fit can be used to calculate a flux-noise level for this $\alpha = 1$ qubit that takes into account both the exponential
nature of non-flux dependent dephasing and the Gaussian nature of
$1/f$ flux-noise decay. Using the same methods outlined in our paper, where we 
specified $\Gamma_{\phi}=2\pi\sqrt{A_{\Phi}|\ln{(2\pi f_{IR}t)}|}D_{\Phi}$, following the approach described in Ref. {[}\citep{ithier2005decoherence_s}{]}, we use the slope of this composite fit
to extract a $1/f$ flux noise level of $A_{\Phi}^{1/2}=1.3~\pm~0.2~\mu \Phi_0$. 
This $\sim10\%$ reduction in the extracted flux-noise level for the $\alpha = 1$ qubit compared to the purely exponential fit {(}$A_{\Phi}^{1/2}~=~1.4~\pm~0.2~\mu \Phi_0${)} brings it closer to the flux-noise level extracted from the fits to the measurements on the $\alpha = 7$ and 4 qubits: $1.3~\pm~0.2~\mu \Phi_0$ and $1.2~\pm~0.2~\mu \Phi_0$, respectively. The Ramsey measurements for these qubits were fit using a purely exponential fit form. It is important to note though, that the $\sim10\%$ reduction in the composite fit extracted flux-noise level for the $\alpha = 1$ qubit is within the errors associated with our flux-noise calculations.

To conclude this fitting study, we have shown that:
\begin{enumerate}
\item The $\alpha = 1$ qubit in this study has a Ramsey decay envelope that is more Gaussian
in nature at high $D_{\Phi}$ values where the dephasing of this qubit is strongly influenced by low-frequency flux noise.
\item Though we have discussed different fitting approaches that better
model the Ramsey decay envelope of qubits influenced by $1/f$ flux-noise,
using a purely exponential decay form for the Ramsey decay simply
puts an upper bound on the extracted flux noise strength. Also, the value of the flux-noise level and the dephasing rates are comparable to those we obtained with the various other fitting approaches.
\item Using a Ramsey fit function that takes into account both the exponential
nature of the $T_{1}$ contribution to the decay envelope and non-flux dependent dephasing, as well as the
Gaussian nature of dephasing due to $1/f$ flux noise, allows us to calculate
a flux noise level for the $\alpha = 1$ qubit that agrees well with the other,
asymmetric qubits on the same sample. This is expected, as qubits of the same geometry on the same chip should experience similar flux noise \citep{sendelbach2008magnetism_s}.
\end{enumerate}

\section{Dephasing Rate Discussion}

\begin{figure}[!b]

\includegraphics[width=1.0\columnwidth]{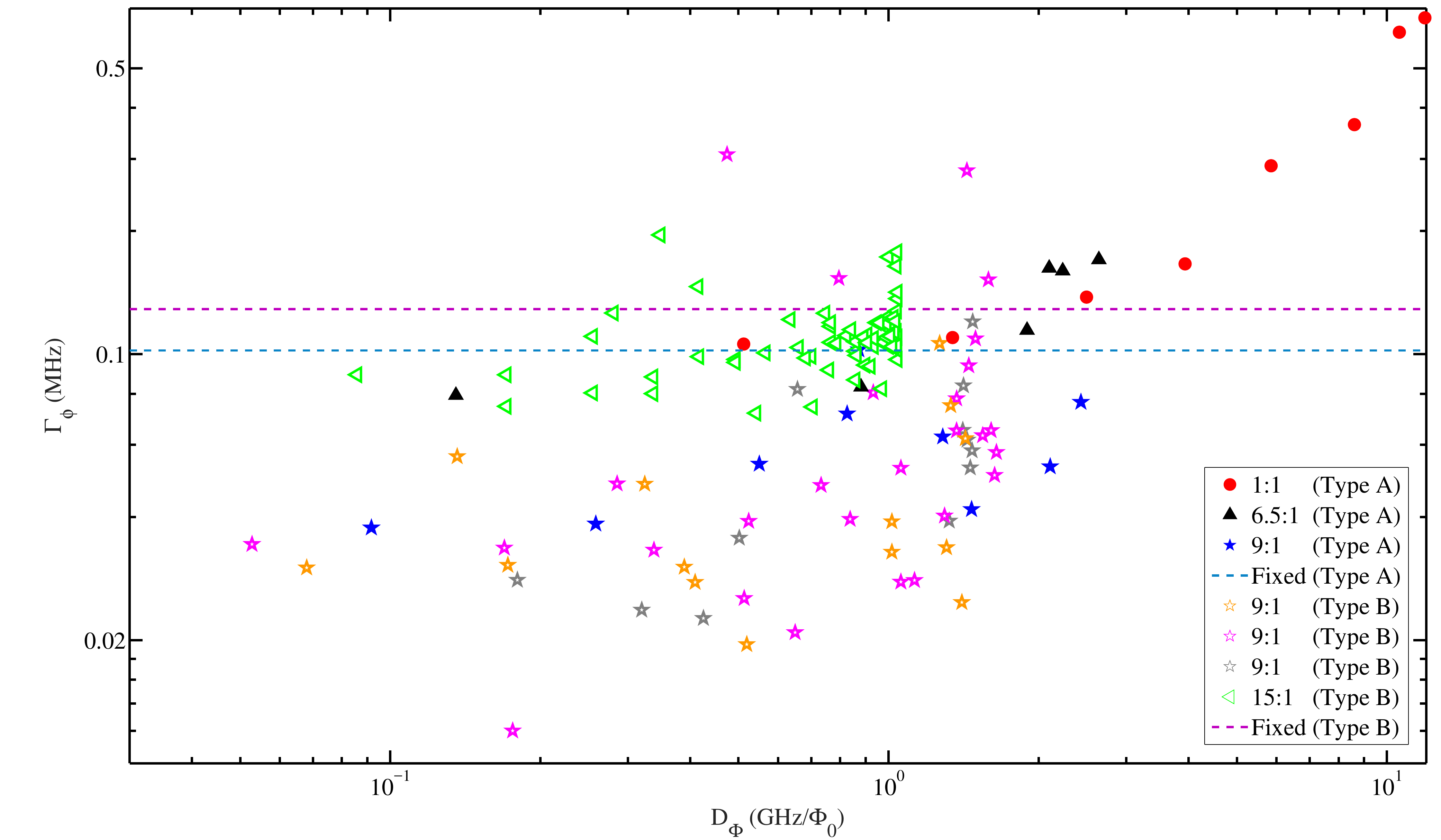}
\caption{$\Gamma_{\phi}$ vs $D_{\Phi}$ for qubits measured during this study that were not included in the main paper. $\Gamma_{\phi}$ for fixed-frequency qubits included as dashed lines. Type A/B qubits were similar in design to those on sample A/B, measured using similar methods and device designs as those described for the corresponding sample type.
\label{fig:9}}
\end{figure}

In Fig. \ref{fig:9} we present dephasing rates for several additional qubits, plotted against $D_{\Phi}$. These qubits were similar to those in our paper, but were prepared on additional chips and measured during additional cools of our cryostats. These data are not included in our paper for reasons of clarity and consistency. However, they are presented here to support the observations found in this study across all qubits measured in both of our labs. 

The first observation we make from Fig. \ref{fig:9} is that a spread in background dephasing rates is measured between both fixed-frequency and tunable qubits. As discussed in our paper, these subtle variations in qubit dephasing rate are not unexpected and are commonly observed in multi-qubit devices \citep{corcoles2015demonstration_s,chow2014implementing_s,Takita2016Dem_s}. While these variations in dephasing rate make the figure somewhat challenging to interpret, we can still draw the same conclusions for this data as those from our main paper. We still observe that the dephasing rate due to flux-noise increases linearly with $D_{\Phi}$ for the lower asymmetry qubits. Again at lower $D_{\Phi}$ values, below $\sim 1\,{\rm GHz}/\Phi_0$, the rate of dephasing is constant within the experimental spread for all qubits. Here, it is important to note that, for several of the qubits shown here and those discussed in our paper, there are specific flux bias points for each qubit where the dephasing rate is anomalously high. These points almost always coincide with places where $T_{1}$ drops sharply at specific frequencies, presumably due to localized coupling to defects in these qubits. Again, this sharp frequency dependence in $T_{1}$ is not unusual for tunable superconducting qubits and is consistent with what others have observed \citep{barends2013coherent_s}.  

The relatively flux-independent dephasing rate at low $D_{\Phi}$ is particularly apparent in the 9:1 qubits we measured. Several of these qubits exhibited the lowest background depahsing rates we observed in our study, between 20 and 40~kHz. These dephasing rates are comparable to current state-of-the-art superconducting qubits \cite{Takita2016Dem_s}. No fixed-frequency qubits were included on the same chips as these 9:1 asymmetric transmons, which prevents us from making a direct comparison with non-flux-noise-driven background dephasing rates as is done in the main paper. Nonetheless, for these 9:1 qubits, we can clearly see that the dephasing rate is essentially flux independent below $\sim 1\,{\rm GHz}/\Phi_0$ even at these low background dephasing levels. This reinforces our statement that asymmetric qubits with a useful level of tunability can be incorporated into future fault-tolerant superconducting qubit devices, significantly aiding scalability in these systems.

\end{document}